\documentclass[preprint,times,tighten]{aastex63}

\usepackage{epsf,color,amsmath}

\usepackage{cancel}

\newcommand{\sfrac}[2]{\mathchoice%
  {\kern0em\raise.5ex\hbox{\the\scriptfont0 #1}\kern-.15em/
    \kern-.15em\lower.25ex\hbox{\the\scriptfont0 #2}}
  {\kern0em\raise.5ex\hbox{\the\scriptfont0 #1}\kern-.15em/
    \kern-.15em\lower.25ex\hbox{\the\scriptfont0 #2}}
  {\kern0em\raise.5ex\hbox{\the\scriptscriptfont0 #1}\kern-.2em/
    \kern-.15em\lower.25ex\hbox{\the\scriptscriptfont0 #2}} {#1\!/#2}}

\newcommand{\enucdot}{\dot{e}_{\rm nuc}}

\newcommand{\C}{$^{12}$C}

\newcommand{\castro}{{\sf Castro}}

\newcommand{\repavg}[1]{\langle #1 \rangle_w}

\newcommand{\gcc}{\mathrm{g~cm^{-3} }}

\newcommand{\half}{\frac{1}{2}}

\def\app#1#2{%
	\mathrel{%
		\setbox0=\hbox{$#1\sim$}%
		\setbox2=\hbox{%
			\rlap{\hbox{$#1\propto$}}%
			\lower1.1\ht0\box0%
		}%
		\raise0.25\ht2\box2%
	}%
}
\def\appropto{\mathpalette\app\relax}

\begin{document}
\title{Dynamics of Laterally Propagating Flames in X-ray Bursts. II. Realistic Burning \& Rotation}

\shorttitle{Lateral Flame Dynamics II}
\shortauthors{Harpole et al.}

\author[0000-0002-1530-781X]{A.\ Harpole}
\affiliation{Dept.\ of Physics and Astronomy, Stony Brook University,
             Stony Brook, NY 11794-3800}

\author[0000-0001-8921-3624]{N.\ M.\ Ford}
\affiliation{Lawrence Berkeley National Laboratory, Berkeley, CA}

\author[0000-0001-6191-4285]{K.\ Eiden}
\affiliation{Dept.\ of Astronomy, University of California, Berkeley,
	CA 94720-3411}
\affiliation{Dept.\ of Physics and Astronomy, Stony Brook University,
	Stony Brook, NY 11794-3800}

\author[0000-0001-8401-030X]{M.\ Zingale}
\affiliation{Dept.\ of Physics and Astronomy, Stony Brook University,
             Stony Brook, NY 11794-3800}

\author[0000-0003-2300-5165]{D.\ E.\ Willcox}
\affiliation{Lawrence Berkeley National Laboratory, Berkeley, CA}

\author[0000-0002-6447-3603]{Y.\ Cavecchi}
\affiliation{Universidad Nacional Aut\'{o}noma de M\'{e}xico, Instituto de Astronom\'{i}a, Ciudad Universitaria, CDMX 04510, Mexico}

\author[0000-0003-0439-4556]{M.\ P.\ Katz}
\affiliation{NVIDIA Corporation}

\correspondingauthor{A.\ Harpole}
\email{alice.harpole@stonybrook.edu}

\begin{abstract}
We continue to investigate two-dimensional laterally propagating flames in type I X-ray 
bursts using fully compressible hydrodynamics simulations.  In the current study we relax 
previous approximations where we artificially boosted the flames. We now use more physically 
realistic reaction rates, thermal conductivities, and rotation rates, exploring the effects 
of neutron star rotation rate and thermal structure on the flame.  We find that at lower 
rotation rates the flame becomes harder to ignite, whereas at higher rotation rates the 
nuclear burning is enhanced by increased confinement from the Coriolis force and the 
flame propagates steadily. At higher crustal temperatures, the flame moves more quickly 
and accelerates as it propagates through the atmosphere. If the temperature is too high, 
instead of a flame propagating across the surface the entire atmosphere burns 
\replaced{steadily}{uniformly}. \added{Our findings could have implications for the 
relationship between observed burst rise times and neutron star rotation and accretion rates.} 
All of the software used for these simulations is freely available.

\end{abstract}


\section{Introduction}\label{Sec:Introduction}

Considerable evidence suggests that ignition in an X-ray burst (XRB) starts
in a localized region and then spreads across the surface of the
neutron star~\citep{bhattacharyya:2007,chakraborty:2014}.  We continue
our study of flame spreading through fully compressible hydrodynamics
simulations of the flame.  Building on our previous
paper~\citep{flame_wave1}, we relax the approximations we made
previously (artificially boosting the speed of the flame in order to
reduce the computational cost) and explore how the flame properties
depend on rotation rate and the thermal structure of the neutron
star. This new set of realistic simulations is possible because of the
work done to offload our simulation code, \castro~\citep{castro_joss}, to GPUs, where it
runs significantly faster.

We investigate the effect of rotation rate on the flame. With the
exception of IGR J17480-2446 (\citealt{altamirano2010discovery}, spinning at $11~\mathrm{Hz}$), \replaced{most observations of
XRBs come from sources with}{most observations of XRBs which come from sources with known rotation rates have} rotation rates of $200-600~\mathrm{Hz}$
\citep{bilous:2019,galloway:2020}. There are a number of
factors that could explain this lack of observations below $200~\mathrm{Hz}$. It
could be that there is some physical process which inhibits the flame
ignition and/or spread at lower rotation rates. It could be that bursts
at lower rotation rates are smaller in amplitude and
therefore more difficult to detect. It could be that it does not have
anything to do with the flame at all, but that neutron stars in
accreting low mass X-ray binaries rarely have rotation rates below $200~\mathrm{Hz}$.
 
Previous studies have found that rotation can have a significant
effect on the flame's propagation. As the rotation rate increases, the
Coriolis force whips the spreading flame up into a hurricane-like
structure \citep{spitkovsky2002,cavecchi:2013}. The stronger Coriolis
force leads to greater confinement of the hot accreted matter, leading
to easier ignition of the flame \citep{art-2015-cavecchi-etal}.

The temperature structure of the accreted fuel layer can also affect
the flame propagation.  \citet{Timmes00} showed that laminar helium flames 
have higher speeds when moving into hotter upstream fuel.
It has been suggested that crustal heating may be stronger at lower
accretion rates and weaker at higher accretion rates, due to the
effect of neutrino losses \citep{Cumming2006,johnston:2019}. On the other hand, at very high accretion rates the atmosphere is so heated that it simmers in place rather than forming a propagating flame \citep{fujimoto1981,bildsten1998thermonuclear,keek2009effect}. 
A shallow heating
mechanism of as yet unknown origin has been found necessary to
reproduce observed properties of XRBs in 1D simulations
\citep{Deibel2015,Turlione2015,Keek2017}.
In our
models, we keep the crust at a constant temperature, so by increasing
this temperature we can effectively increase the crustal heating, shallow heating
and/or
mimic the effects of accretion-induced heating. 

In the following sections, we conduct a series of simulations at various rotation 
rates and crustal temperatures to investigate their effects on the flame. We find that 
at lower rotation rates, the flame itself becomes harder to ignite. At higher 
rotation rates, nuclear burning is enhanced and the flame propagates steadily. 
At higher crustal temperatures, burning is greatly enhanced and the flame accelerates as 
it propagates. We discuss the implications that this may have for burst physics 
and observations.

\section{Numerical Approach}\label{Sec:numerics}

We use the \castro\ hydrodynamics code \citep{castro,castro_joss} and
the simulation framework introduced in \citet{flame_wave1}.  The
current simulations are all performed in a two-dimensional
axisymmetric geometry.  For these axisymmetric simulations, we add an
additional geometric source term from \citet{bernard-champmartin} that
captures the effects of the divergence of the flux operating on the
azimuthal unit vector.  This term is a small correction, but was
missing from our previous simulations.  The simulation framework
initializes a fuel layer in hydrostatic equilibrium, laterally
blending a hot model on the left side of the domain (the coordinate
origin) and a cool model on the right.  The initial temperature
gradient between the hot and cool fluids drives a laterally
propagating flame through the cool fuel.  In our original set of
calculations \citep{flame_wave1}, in order to make the 
simulations computationally feasible we artificially boosted the flame speed by adjusting the
conductivity and reaction rate to produce a flame moving 5--10$\times$
faster than the nominal laminar flame speed.  We also used high
rotation rates ($\geq 1000~\mathrm{Hz}$) to reduce the lateral lengthscale at which the Coriolis
force balances the lateral flame spreading in order to reduce the size 
of the simulation domain. The port of \castro\ to GPUs~\citep{sc20_gpu} 
significantly improved its overall performance, enabling
us to run these new simulations without the previous approximations while
continuing to resolve the burning front. For these
simulations, we no longer boost the flame speed---the true
conductivities (taken from \citealt{Timmes00}) and reaction rates are
used. We are also able to use slower, more physically realistic rotation rates.
  We continue to use a 13-isotope $\alpha$-chain to describe the
helium burning.  

The initial model is set up in the same fashion as described in
\citet{flame_wave1}.  In particular, we create a ``hot'' and ``cool''
hydrostatic model representing the ash and fuel states and blend the
two models laterally to create a hot region near the origin of the
coordinates and a smooth transition to the cooler region at larger
radii. \added{This hot region extends out to $r_\mathrm{pert} =4.096\times 10^4 \mathrm{cm}$.}
 The cool initial model is characterized by three temperatures:
$T_\mathrm{star}$ is the isothermal temperature of the underlying
neutron star, $T_\mathrm{hi}$ is the temperature at the base of the
fuel layer, and $T_\mathrm{lo}$ is the minimum temperature of the
atmosphere. The atmosphere structure is isentropic as it falls from
$T_\mathrm{hi}$ down to $T_\mathrm{lo}$, \added{and so is marginally 
unstable to convection}.  For the hot model, we
replace $T_\mathrm{hi}$ with $T_\mathrm{hi} + \delta T$.  In the
calculations presented here, we explore the structure of the initial
models by varying these parameters.  All models have the same
peak temperature in the hot model, $T_\mathrm{hi} + \delta T$.

\begin{figure}[t]
    \centering
    \plotone{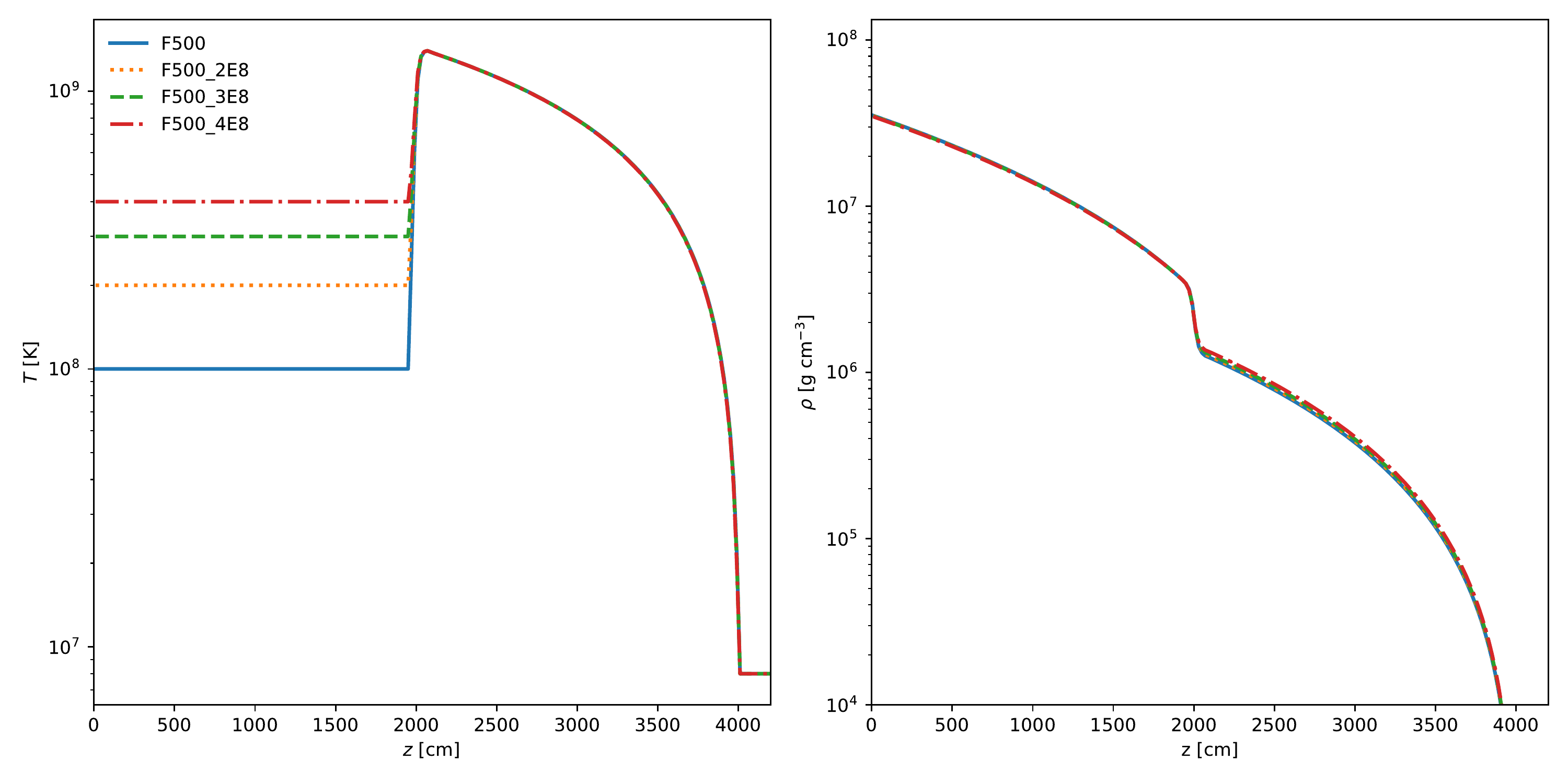}
    \caption{\label{fig:hot_profiles} Initial temperature structure (left panel) and density
    structure (right panel) as a function of height in the ``hot'' state.}
\end{figure}

\begin{figure}[t]
    \centering
    \plotone{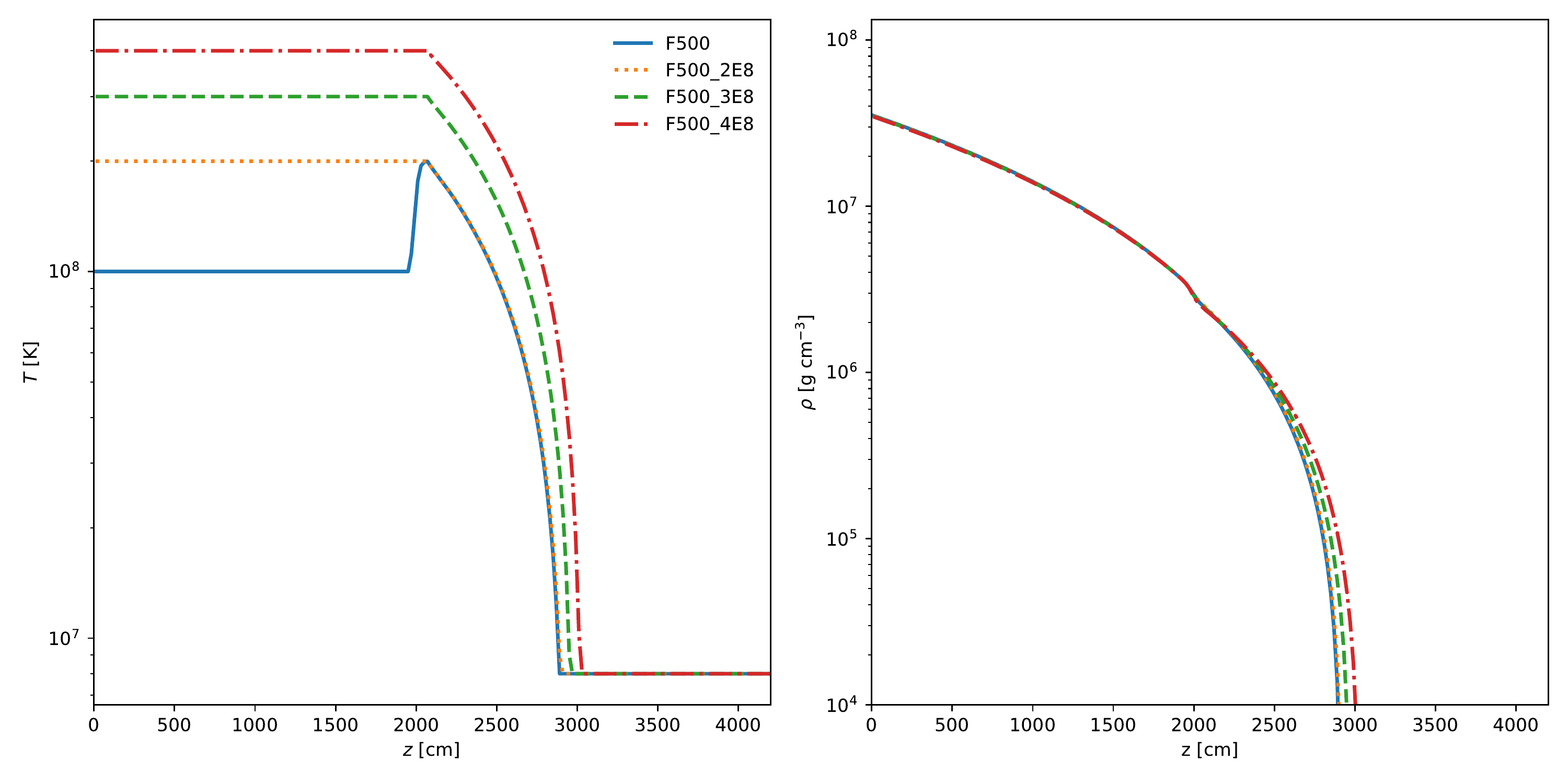}
    \caption{\label{fig:cool_profiles} Initial temperature structure (left panel) and density
    structure (right panel) as a function of height in the ``cool'' state.}
\end{figure}

For the current simulations, we explore a variety of initial rotation
rate and temperature conditions for the flame. The main parameters
describing the models and the names by which we shall refer to them in this paper 
are provided in Table~\ref{table:runs}. Figure~\ref{fig:hot_profiles} shows the
temperature and density structure for our hot models and
Figure~\ref{fig:cool_profiles} shows the temperature and density
structure for the cool models.

\begin{deluxetable}{lrrrrr}
	\tablecaption{\label{table:runs} Rotation rate and temperature properties of the simulations. In the left-hand column we list the names we shall use to refer to each simulation throughout this paper.}
	\tablehead{\colhead{run} & \colhead{Rotation Rate (Hz)} & \colhead{$\delta T$ (K)} & \colhead{$T_\mathrm{hi}$ (K)} & \colhead{$T_\mathrm{star}$ (K)} & \colhead{$T_\mathrm{lo}$ (K)}} 
	\startdata
	{\tt F1000}     & $1000$ & $1.2\times 10^9$ & $2\times 10^8$ & $10^8$ & $8\times 10^6$ \\
	{\tt F500}      & $500$ & $1.2\times 10^9$ & $2\times 10^8$ & $10^8$ & $8\times 10^6$ \\
	{\tt F500\_2E8} & $500$ & $1.2\times 10^9$ & $2\times 10^8$ & $2\times 10^8$ & $8\times 10^6$ \\
	{\tt F500\_3E8} & $500$ & $1.1\times 10^9$ & $3\times 10^8$ & $3\times 10^8$ & $8\times 10^6$ \\
	{\tt F500\_4E8} & $500$ & $10^9$ & $4\times 10^8$ & $4\times 10^8$ & $8\times 10^6$ \\
	{\tt F250}      & $250$ & $1.2\times 10^9$ & $2\times 10^8$ & $10^8$ & $8\times 10^6$ \\
	\enddata
\end{deluxetable}

\section{Simulations and Results}\label{Sec:results}

We present six simulations in total, summarized in
Table~\ref{table:runs}.  These simulations encompass three different
rotation rates: $250~\mathrm{Hz}$, $500~\mathrm{Hz}$, and $1000~\mathrm{Hz}$, 
and for the $500~\mathrm{Hz}$ run, four
different temperature profiles.  In the following subsections, we look
at how the flame properties depend on the model parameters.  All
simulations are run in a domain of $1.8432\times 10^5~\mathrm{cm} \times
3.072\times 10^4~\mathrm{cm}$ with a coarse grid of $1152 \times 192$
zones and two levels of refinement (the first level refining the resolution 
by factor of four, and the second by a factor of two again).  This gives a
fine-grid resolution of $20~\mathrm{cm}$.  In these simulations, refinement 
is carried out in all zones within the atmosphere
with density $\rho > 2.5\times 10^4~\gcc$. We use an axisymmetric coordinate system, 
with the horizontal $r$-direction pointing along the surface of the star and the 
vertical $z$-direction pointing perpendicular to the surface.

\begin{figure}
	\plotone{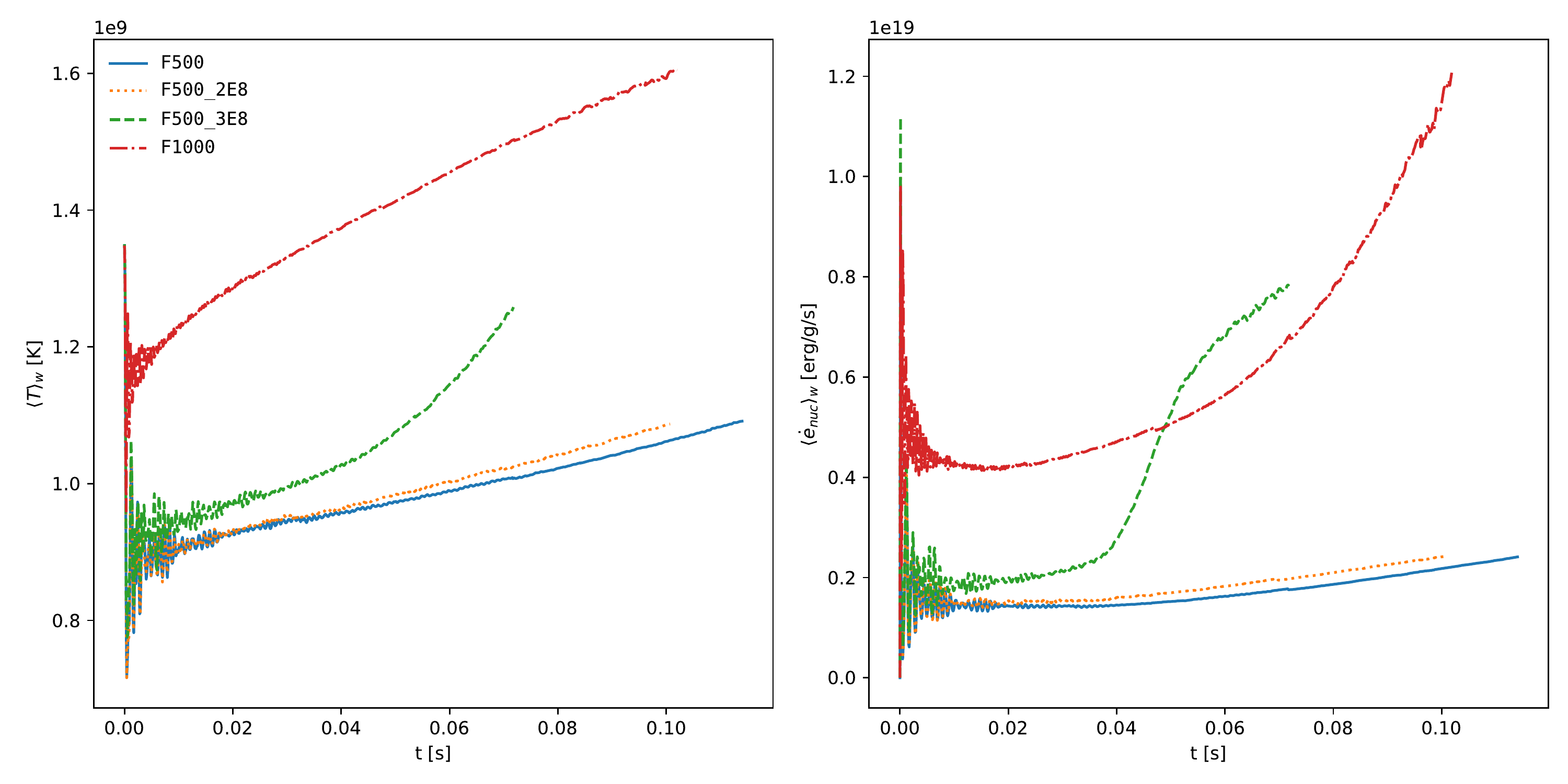}
	\caption{\label{fig:maxima}Estimates of temperature (left panel) and nuclear energy generation 
	rate (right panel) in the burning region as functions of time. The quantities on the vertical 
	axes are the mass-weighted averages defined in Equation~\ref{eqn:top_percentile_avg}.}
\end{figure}

For some of our analysis, we would like to have a means of estimating the temperature ($T$) and 
nuclear energy generation rate ($\enucdot$) in the burning region of each simulation. For this 
purpose, we define the mass-weighted average $\repavg{Q}$ of some quantity $Q$ to be
\begin{equation}
	\label{eqn:top_percentile_avg}
	\repavg{Q} \equiv \frac{\sum_{c_i} m(c_i) Q(c_i)}{\sum_{c_i} m(c_i)}; ~c_i \in C_{99}(Q).
\end{equation}
\noindent Here, $C_{99}(Q)$ is the set of grid cells with $Q$ values in the top percentile, 
$Q(c_i)$ is the value of $Q$ in cell $c_i$, and $m(c_i)$ is the total mass in cell $c_i$. 
Using $\repavg{Q}$ instead of simply taking the maximum of the quantity across the entire 
simulation domain allows us to track changes over the domain as a whole rather than at a single 
localized point. This will therefore be a better reflection of the overall behavior of the flame 
rather than of a single localized fluctuation. Figure \ref{fig:maxima} shows $\repavg{T}$ and 
$\repavg{\enucdot}$ as functions of time for the subset of our runs that achieve a propagating 
flame. This figure is referenced throughout the subsequent sections.

\subsection{Effect of Rotation Rate on Flame Structure}\label{ssec:rot_structure}

\begin{figure}[t]
\centering
\plotone{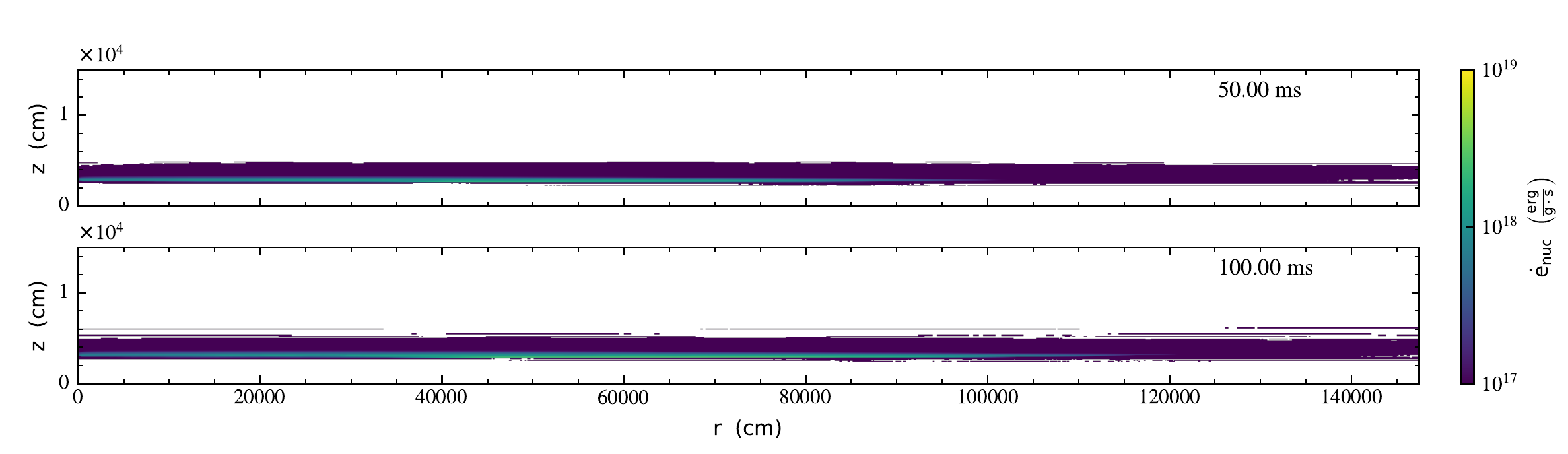}
\caption{\label{fig:time_series_enuc_500} Time series of the $500~\mathrm{Hz}$ run {\tt F500} showing the nuclear energy generation rate, $\dot{e}_\mathrm{nuc}$.}
\end{figure}

\begin{figure}[t]
\centering
\plotone{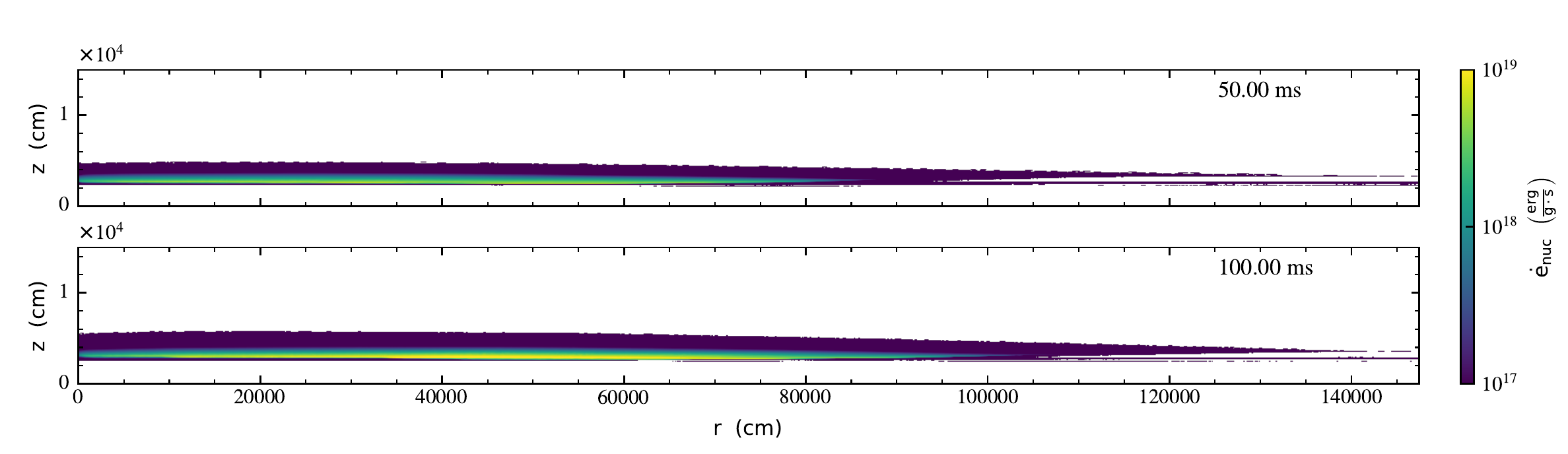}
\caption{\label{fig:time_series_enuc_1000} Time series of the $1000~\mathrm{Hz}$ run {\tt F1000} showing the nuclear energy generation rate, $\dot{e}_\mathrm{nuc}$.}
\end{figure}

We run three models ({\tt F250}, {\tt F500}, and {\tt F1000}) with the same initial model in terms 
of temperature but differing rotation rates. We saw in \citet{flame_wave1} that 
increasing the neutron star rotation rate reduces the horizontal lengthscale of the flame. An 
estimate of this lengthscale is given by the Rossby radius of deformation, $L_R$. The Rossby radius 
may be thought of as the scale over which the balance between the Coriolis force and horizontal 
pressure gradient becomes important, and is approximately given by
\begin{equation}
	\label{eqn:rossby}
	L_R \approx \frac{\sqrt{g H_0}}{\Omega},
\end{equation}
where $g$ is the gravitational acceleration, $H_0$ is the atmospheric scale height, and $\Omega$ is the 
neutron star rotation rate. \added{If we take $g \sim 1.5 \times 10^{14}~\mathrm{cm}~\mathrm{s}^{-2}$ and $H_0\sim 10^3~\mathrm{cm}$, then for the $500~\mathrm{Hz}$ system we find the Rossby length to approximately be $L_R \sim 1.2 \times 10^5~\mathrm{cm}$. This estimate is likely to be an overestimate of the true value of the Rossby length in our simulations, as we saw that the flame is confined on a much smaller length scale.} In Figure \ref{fig:time_series_enuc_500} and Figure 
\ref{fig:time_series_enuc_1000}, we use $\dot{e}_\mathrm{nuc}$ measured at $50~\mathrm{ms}$ and $100~\mathrm{ms}$ to discern the horizontal extent of the flame at different rotation rates. Taking the edge at greatest radius of the bright teal/green region where the most significant energy generation is occurring as the leading edge of the flame in each plot, we see that the horizontal extent of the $1000~\mathrm{Hz}$ flame ($\tt{F1000}$) appears to be 
reduced compared to the lower rotation $500~\mathrm{Hz}$ run ($\tt{F500}$). \added{As the natural aspect ratio makes it hard to see the flame structure, we show the energy generation rate for the $\tt{F1000}$ simulation with the vertical extent stretched in Figure~\ref{fig:flame_stretch}.}  From Equation 
\ref{eqn:rossby}, we can see that increasing the rotation rate from $500~\mathrm{Hz}$ to 
$1000~\mathrm{Hz}$ should decrease $L_R$ by a factor of two, and that the greater confinement from 
the Coriolis force should reduce the horizontal extent of the flame by a similar factor. However, 
the Rossby radius is only an approximate measure of this horizontal lengthscale, and in our 
simulations we see that this scaling does not work so well for all rotation rates. The simulations 
seem to follow the theoretical scaling described in Equation \ref{eqn:rossby} more closely at 
higher rotation rates ($1000~\mathrm{Hz}$ and higher), based on the results of \citet{flame_wave1}.

\added{
\begin{figure}[t]
\centering
\plotone{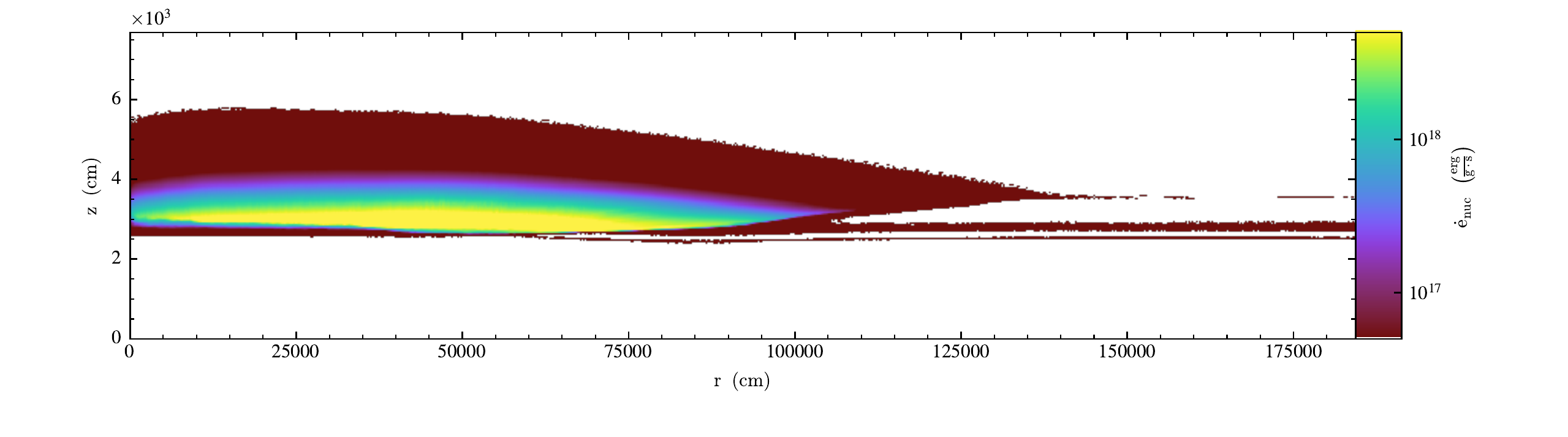}
\caption{\label{fig:flame_stretch} Energy generation rate for run {\tt F1000} at 100 ms with the vertical extent stretched to show detail.}
\end{figure}
}

\begin{figure}[t]
\centering
\plotone{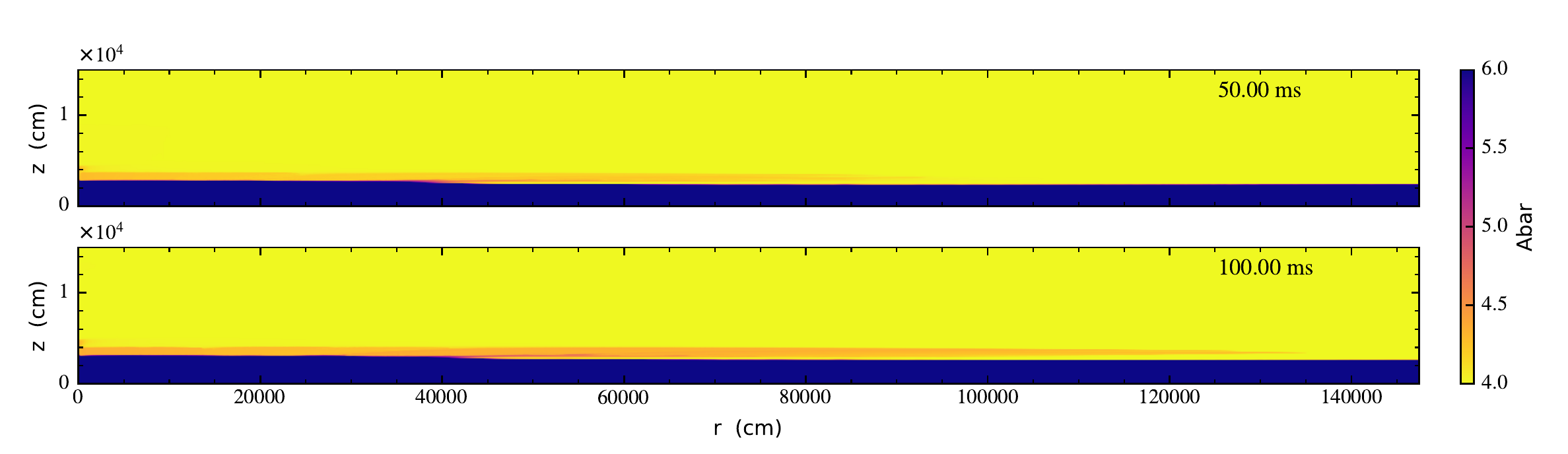}
\caption{\label{fig:time_series_500} Time series of the $500~\mathrm{Hz}$ run {\tt F500} showing the mean molecular weight, $\bar{A}$.}
\end{figure}

\begin{figure}[t]
\centering
\plotone{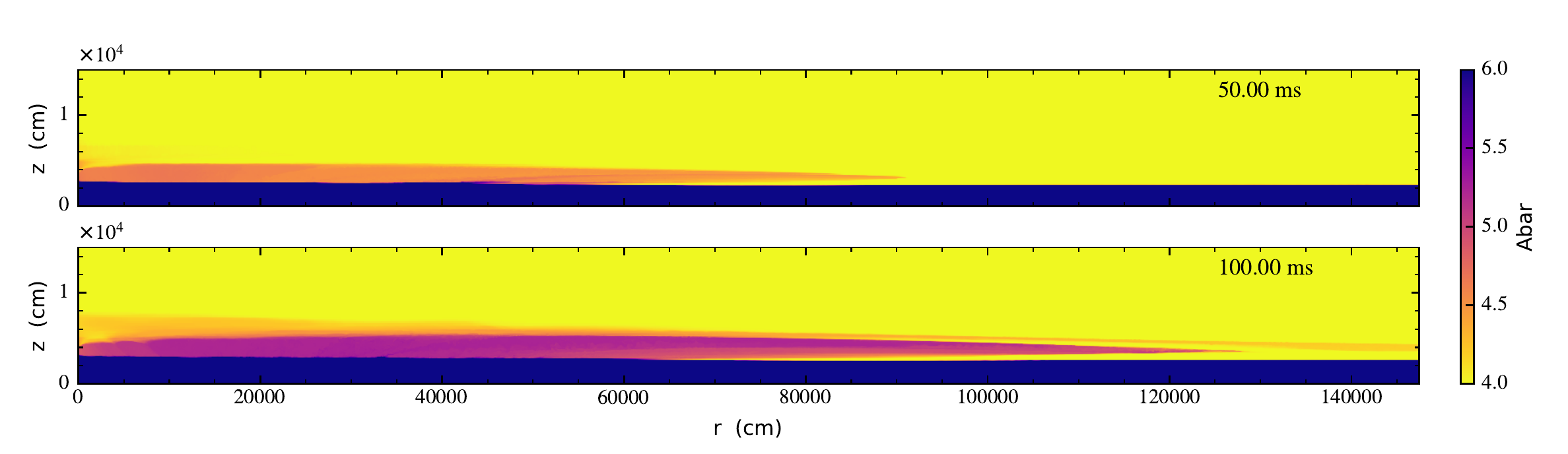}
\caption{\label{fig:time_series_1000} Time series of the $1000~\mathrm{Hz}$ run {\tt F1000} showing the mean molecular weight, $\bar{A}$.}
\end{figure}

The {\tt F500} and {\tt F1000} runs both qualitatively resemble the flame structure in 
\citet{flame_wave1} --- a laterally propagating flame that is lifted off of the bottom of the fuel 
layer --- but they differ in their burning structures. Figures~\ref{fig:time_series_500} and 
\ref{fig:time_series_1000} show time series of the mean molecular weight, $\bar{A}$, for the {\tt 
F500} and {\tt F1000} runs.  Compared to those in \cite{flame_wave1}, ashes behind the flame do not 
reach as high atomic weights. This is not surprising, since those previous runs artificially 
boosted the reaction rates.  Comparing these two new runs, the burning is much more evolved for the 
higher rotation rate, and the ash is actually able to move ahead of the flame front (visible in the 
Figure  \ref{fig:time_series_1000} $100~\mathrm{ms}$ snapshot). We believe that this is because the increased 
rotation better confines the initial perturbation and subsequent expansion from the burning, 
increasing the temperature and density in the flame front such that the reaction rate increases, 
which allows the reactions to progress further. The $\dot{e}_\mathrm{nuc}$ plots in Figure 
\ref{fig:time_series_enuc_1000} also support this interpretation, with the region of the 
flame front nearest to the crust in the {\tt F1000} run reaching higher $\dot{e}_\mathrm{nuc}$ values 
than for the {\tt F500} run in Figure~\ref{fig:time_series_enuc_500}. In contrast to {\tt F500} 
and {\tt F1000}, the lowest rotation run --- {\tt F250} --- failed to ignite. The lack of ignition 
for {\tt F250} also aligns with the reasoning given above, with the lower rotation in this case 
potentially leading to insufficient confinement such that the temperature and density required for 
ignition is not achieved. In this scenario, another source of confinement (e.g.\ magnetic fields, see \cite{art-2016-cavecchi-etal}) 
would need to take over at lower rotation rates to allow a burst to occur, at least for the initial 
model used here. Given that the size of our domain is $\sim 1~L_R$ for {\tt F250} (using Equation~\ref{eqn:rossby}), it is also possible that we simply cannot confine the flame sufficiently with our current domain width. We see in Figure~\ref{fig:flame_speeds_1} (discussed further in Section~\ref{ssec:rot_propagation}) that the {\tt F500} flame took longer to achieve steady propagation than the {\tt F1000} flame. It may therefore also be that we did not run our simulation for long enough to see the {\tt F250} flame achieve the conditions required for ignition and steady propagation.

\begin{figure}[t]
    \centering
    \plotone{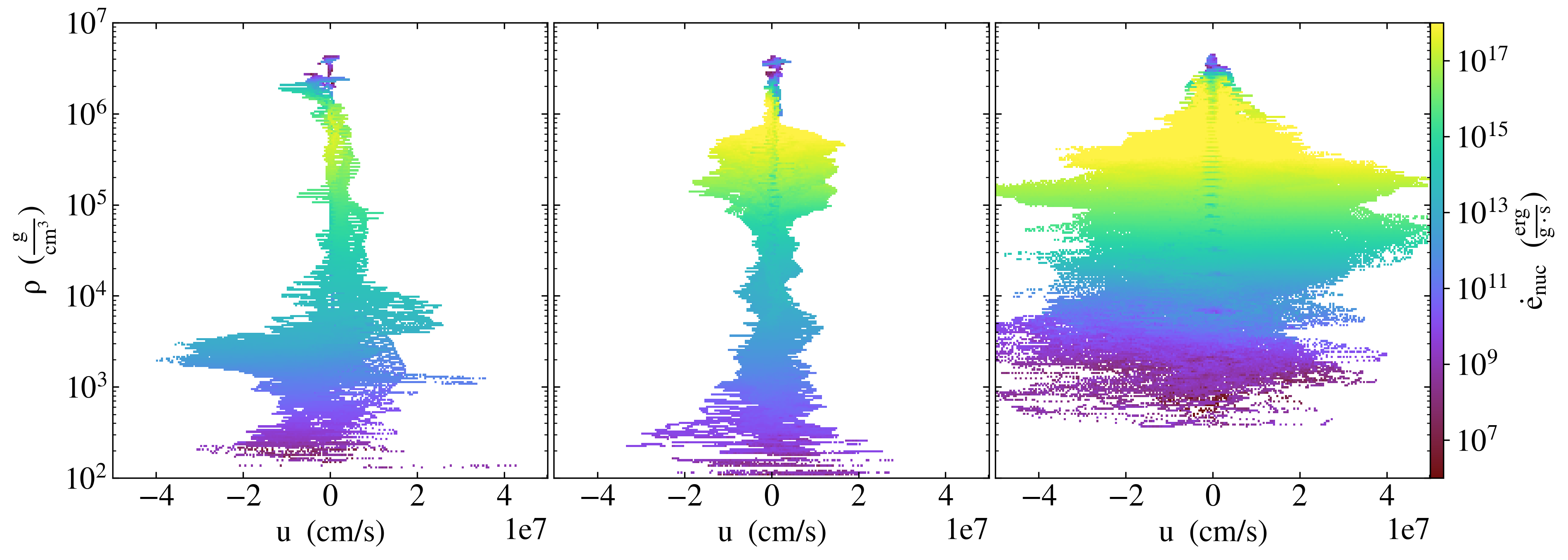}{}
    \caption{\label{fig:urho}Phase plot of the horizontal $r$-velocity $u$, density $\rho$ and the nuclear energy generation rate $\dot{e}_{\rm nuc}$ for the 250, 500 and $1000~\mathrm{Hz}$ runs ({\tt F250}, {\tt F500} and {\tt F1000}) at $t = 100~\mathrm{ms}$. The slightly lower $\dot{e}_{\rm nuc}$ values along the $u = 0$ axis are most likely a numerical artifact related to the density gradient setup and finite resolution of these simulations.}
\end{figure}

\begin{figure}[t]
    \centering
    \plotone{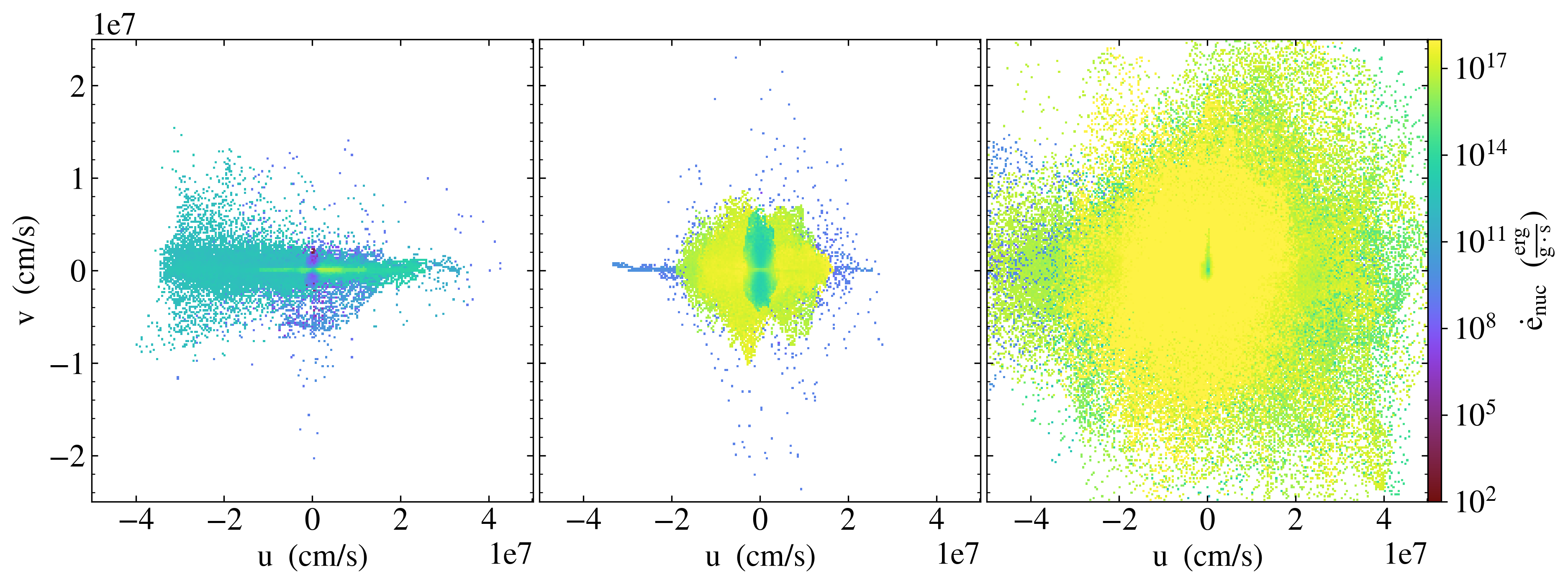}
    \caption{\label{fig:uv}Phase plot of the horizontal $r$-velocity $u$, vertical $z$-velocity $v$ and the nuclear energy generation rate $\dot{e}_{\rm nuc}$ for the 250, 500 and $1000~\mathrm{Hz}$ runs ({\tt F250}, {\tt F500} and {\tt F1000}) at $t = 100~\mathrm{ms}$.}
\end{figure}

Burning in the {\tt F500} and {\tt F1000} runs is concentrated in a dense region with circular 
motion. In Figure~\ref{fig:urho}, which compares the horizontal $r$-velocity $u$, density $\rho$ and the 
nuclear energy generation rate $\dot{e}_{\rm nuc}$ for the {\tt F250}, {\tt F500}, and {\tt F1000} 
runs, most of the burning for each of the simulations occurs in a high density region $\rho > 3 
\times 10^5~{\rm g\,cm^{-3}}$. The fluid in this dense, high energy generation region undergoes 
vortical motion, shown in the Figure~\ref{fig:uv} phase plots comparing $u$, the vertical $z$-velocity $v$ and $\dot{e}_{\rm 
nuc}$. This most likely corresponds to the leading edge of the flame where fresh fuel is being entrained.
This feature is not developed in the $250~\mathrm{Hz}$ flame in Figure~\ref{fig:uv} (left panel); it could potentially develop 
at later times (past the point at which we terminated our simulation), or the burning could just 
fizzle out and the flame fail to ignite entirely. 

\begin{figure}[t]
    \centering
    \plotone{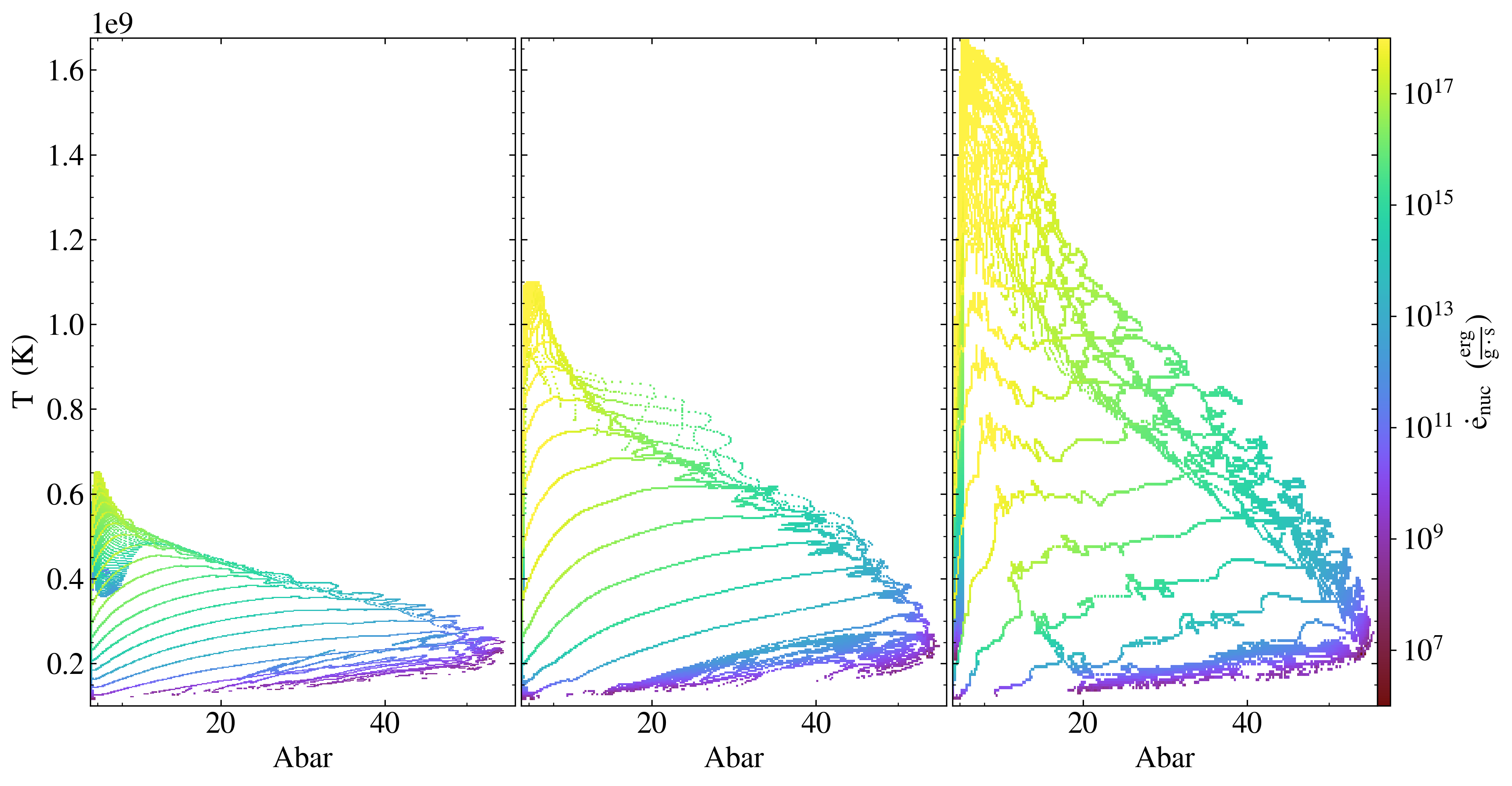}
    \caption{\label{fig:abar}Phase plot of the mean molecular weight $\bar{A}$, temperature $T$ and the nuclear energy generation rate $\dot{e}_{\rm nuc}$ for the 250, 500 and $1000~\mathrm{Hz}$ runs ({\tt F250}, {\tt F500} and {\tt F1000}) at $t = 100~\mathrm{ms}$.}
\end{figure}

The mean molecular weight $\bar{A}$ within each of our simulations seems to grow along defined tracks confined to certain temperatures $T$, as shown in the Figure \ref{fig:abar} phase plots. We believe that the tracks in the plot correspond to different burning trajectories in phase space resulting from different thermodynamic conditions. Comparing Figure \ref{fig:abar} to \citet{flame_wave1}, these tracks are much more neat and clearly defined. The ``messiness" of the tracks may be dependent on how mixed the flame interior is. Since these new simulations are un-boosted, they may be inherently less mixed than those in \citet{flame_wave1}. {\tt F1000} aligns with this interpretation: its $\bar{A}$ tracks are somewhat disrupted compared to the slower rotation runs, possibly due to the more vigorous mixing of the vortex at the flame front. Comparing the different runs, we also see that as the rotation rate increases, so does the peak temperature. This makes sense if higher rotation leads to a more concentrated, intense vortex near the flame front. It also agrees with our earlier interpretation of the enhanced burning seen in Figure \ref{fig:time_series_1000} for {\tt F1000}. 

\subsection{Effect of Rotation Rate on Flame Propagation}\label{ssec:rot_propagation}

For the purpose of measuring the flame propagation speed and acceleration, we track the position of 
each of our flames as a function of time. We define the position in terms of a specific value of 
the energy generation rate, $\enucdot$, as we did in \citet{flame_wave1}. To recapitulate: we first 
reduce the 2D $\enucdot$ data for each simulation run to a set of 1D radial profiles by averaging
over the vertical coordinate. After averaging, we take our reference $\enucdot$ value to be some 
fraction of the global $\enucdot$ maximum across all of these profiles. Since the flames in our 
simulations propagate in the positive horizontal direction, we then search the region of each profile 
at greater radius than the local $\enucdot$ maximum for the point where the $\enucdot$ first drops 
below this reference value. This point gives us the location of our flame front.

In \citet{flame_wave1}, we used $0.1\%$ of the global $\enucdot$ maximum for our reference value. 
For the high temperature unboosted flames, however, we found that the $\enucdot$ profiles failed to 
reach that small a value across the domain at most times, which prevented us from obtaining reliable 
position measurements. We therefore use $1\%$ of the global $\enucdot$ maximum in this paper
rather than $0.1\%$. This is still sufficiently small that our measurements are not overly 
sensitive to turbulence and other local fluid motions (the issue with simply tracking the local 
maximum), but allows us to avoid the pitfall encountered by the $0.1\%$ metric.

\begin{deluxetable}{lrrr}
	\tablecaption{\label{table:flame_speeds} Flame speed and acceleration values measured for each 
	simulation. $v_0$ is the flame velocity in the $+r$ direction at $t 
	= 0~\mathrm{ms}$ and $a$ is the acceleration of the flame. The initial flame 
	velocities and accelerations are derived from a quadratic least-squares fit to each of the 
	datasets for times $t \gtrsim 20~\mathrm{ms}$. Using these fit parameters we calculate the 
	velocity at $t = 50~\mathrm{ms}$, $v_{50}$, at which point the flames have reached steady 
	propagation.}
	\tablehead{\colhead{run} & \colhead{$v_0$ (km s$^{-1}$)} & \colhead{$a$ (km s$^{-2}$)} & \colhead{$v_{50}$ (km s$^{-1}$)}} 
	\startdata
  {\tt F1000}     & $3.414 \pm 0.008$ & $-1.03 \pm 0.13$ & $3.36 \pm 0.01$\\
  {\tt F500}      & $3.077 \pm 0.013$ & $5.41 \pm 0.19$ & $3.35 \pm 0.02$\\
  {\tt F500\_2E8} & $3.760 \pm 0.015$ & $1.74 \pm 0.25$ & $3.85 \pm 0.02$\\
  {\tt F500\_3E8} & $-5.293 \pm 0.100$ & $357.41 \pm 2.15$ & $12.58 \pm 0.15$\\
	\enddata
\end{deluxetable}

\begin{figure}[t]
	\centering
	\plotone{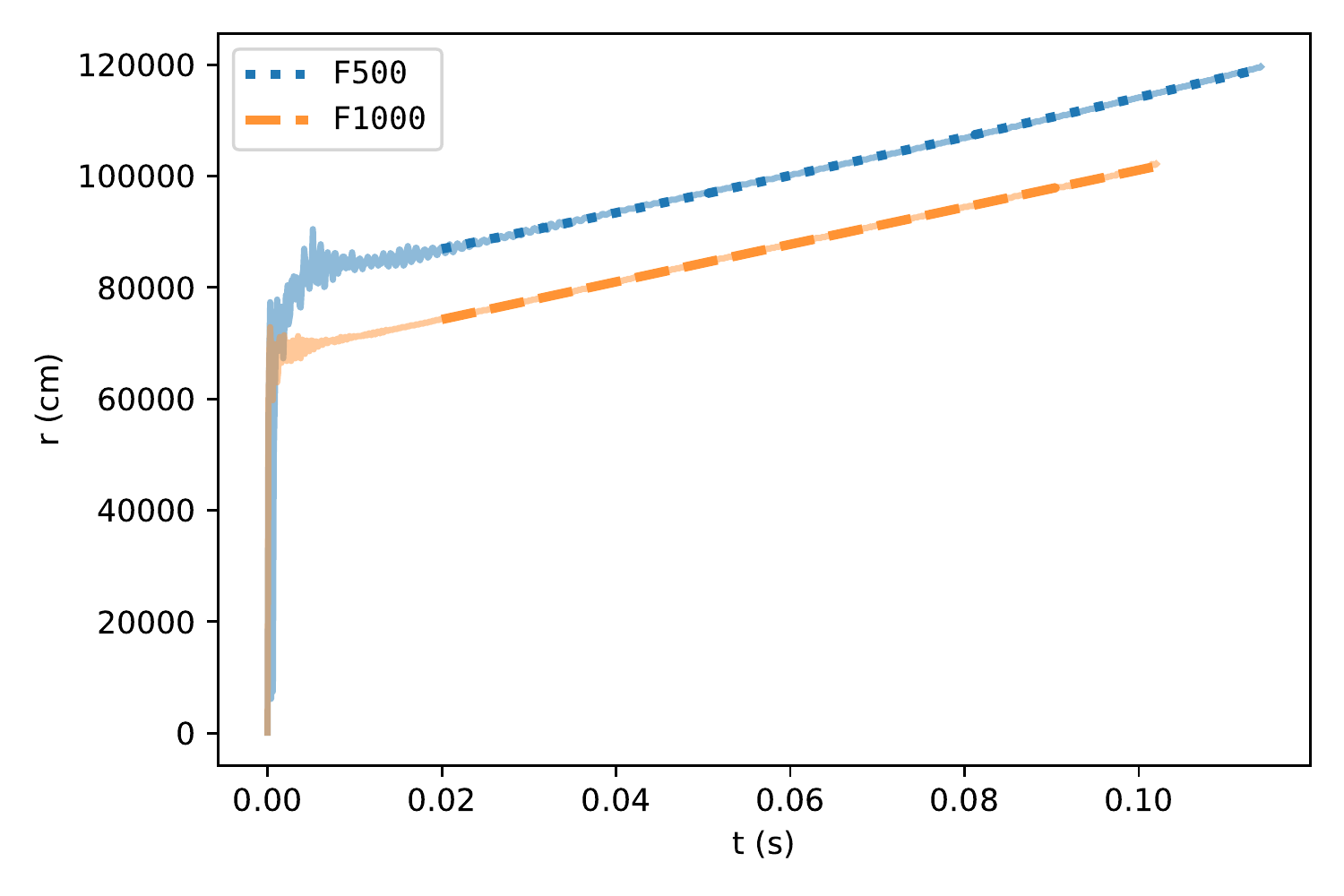}
	\caption{\label{fig:flame_speeds_1} Flame front position vs.\ time for the standard ($10^8~\mathrm{K}$) $500~\mathrm{Hz}$ and $1000~\mathrm{Hz}$ runs ({\tt F500} and {\tt F1000}). The dashed lines
		show quadratic least-squares fits to the data for $t \gtrsim 20~\mathrm{ms}$.}
\end{figure}

Figure~\ref{fig:flame_speeds_1} gives the radial position of the flame front as a function of time 
for the $\tt{F500}$ and $\tt{F1000}$ runs (blue and orange, respectively) to show the dependence on 
rotation rate. In \citet{flame_wave1}, we applied a linear least-squares fit to the flame front 
position as a function of time to estimate the propagation velocity. As some of the flames in this 
set of simulation runs exhibit significant acceleration, for this study we instead fit the data 
with a quadratic curve of the form
\begin{equation}
	\label{eqn:quadratic_fit}
	r(t) = \half a t^2 + v_0 t + r_0,
\end{equation}
\noindent where the parameter $a$ is the acceleration of the flame, $v_0$ is the velocity at $t = 
0~\mathrm{ms}$, and $r_0$ is the flame front position at $t = 0~\mathrm{ms}$. We do not include the data points 
with $t \lesssim 20~\mathrm{ms}$ when performing the fit, since these correspond to the initial 
transient period before the flame has begun to propagate steadily. The values of $a$ and $v_0$ for 
the full suite of simulation runs are provided in Table~\ref{table:flame_speeds}. Note that $v_0$ 
is only a parameter that may be used to calculate the flame speed at an arbitrary time. It is not 
an estimate of the true initial velocity of the flame, since the flame has not achieved ignition 
yet at $t = 0~\mathrm{ms}$. We use the fit parameters to calculate the flame speeds at $t = 
50~\mathrm{ms}$ (when the flame has reached steady propagation), given in the fourth column of 
Table \ref{table:flame_speeds}.

\added{For the simple case of a laminar flame front driven by conduction, the propagation rate, 
$s_l$, of the flame is approximately
\begin{equation}
	\label{eqn:laminar_flame_speed}
	s_l \approx \left( \frac{\mathcal{D} \enucdot}{E} \right)^{\half}.
\end{equation}
Here, $\mathcal{D} = k_{th} / (\rho c_v)$ is the thermal diffusivity with thermal conductivity $k_{th}$ and specific heat capacity $c_v$, $\enucdot$ is the specific
nuclear energy generation rate, and $E$ is the specific total energy \citep{timmeswoosley:1992}. 
\citet{cavecchi:2013} explored the effects of rotation on the flame's propagation, and noted that the 
flame speed in their simulations scaled with the ratio $L_R/H_0$ of Rossby length to atmospheric 
scale height. Combining this scaling with Equation \ref{eqn:laminar_flame_speed} yields a revised 
estimate for the flame speed, $s$:
\begin{equation}
	s \approx \frac{L_R}{H_0} s_l \approx \frac{L_R}{H_0} \left( \frac{\mathcal{D} \enucdot}{E} 
	\right)^{\half}.
\end{equation}
Observing that $L_R$ is inversely proportional to the angular rotation frequency $\Omega$ and 
and neglecting variations in $H_0$ and $E$, we can then derive the following approximate scaling 
relation:
\begin{equation}
	\label{eqn:flame_speed_scaling}
	s \appropto \frac{(\mathcal{D} \enucdot)^{\half}}{\Omega}.
\end{equation}
We were able to test the validity of this scaling relation in \citet{flame_wave1}, where we studied 
the effects of varying the thermal conductivity/diffusivity, energy generation rate, and rotation 
rate. These results supported the findings of \citet{cavecchi:2013}. Despite this formulation of 
the flame speed neglecting some relevant physics (e.g.\ turbulence), the relation proved
consistent with our simulation results.}

\deleted{We found in \citet{flame_wave1} that the flame speed, $s$, obeys the following relation:
$s \appropto L_R (\enucdot)^{\half} \propto \frac{(\enucdot)^{\half}}{\Omega},$
where $\enucdot$ is the specific nuclear energy generation rate and $L_R$ is the Rossby length 
described by Equation \ref{eqn:rossby}. This finding was consistent with the results of 
\citet{cavecchi:2013}, who noted that the flame speeds in their simulations scaled with the Rossby 
length.}

\added{This study explores a more physical parameter space than \citet{flame_wave1}}, and as seen 
in 
Figure~\ref{fig:flame_speeds_1}, there is no clear inverse scaling of the 
flame speed with rotation rate in the new set of runs. We observed earlier, however, that nuclear 
reactions progress more quickly at higher rotation rate. This results in a higher 
$\enucdot$ --- up to three to four times higher near the burning front after the flame ignites (see 
Figure \ref{fig:maxima}) --- which may counteract the reduction in flame speed from the increased 
Coriolis confinement. Comparing accelerations, we also observe that $\tt{F500}$ accelerates faster 
than $\tt{F1000}$, which appears to experience a small deceleration at early times. This disparity 
may be a direct result of the difference in Coriolis force.


\subsection{Effect of Temperature on Flame Structure}\label{ssec:temp_structure}

To explore the effect of different initial temperature configurations, we run four simulations fixed at a rotation rate of $500~\mathrm{Hz}$ with temperatures as shown in Table~\ref{table:runs}. For all the $500~\mathrm{Hz}$ simulations (with the exception of the coolest run, $\tt{F500}$), we set $T_\mathrm{star}$ = $T_\mathrm{hi}$, scaling $\delta T$ accordingly to maintain a consistent value of $T_\mathrm{hi} + \delta T$. If we let $T_\mathrm{star} < T_\mathrm{hi}$, the cooler neutron star surface might act as a heat sink and siphon away energy that would otherwise go into heating the burning layer. By choosing $T_\mathrm{star}$ = $T_\mathrm{hi}$, we can effectively explore simulations with greater surface heating. There are several physically distinct mechanisms that could produce an increased temperature at the crust: crustal heating, some other form of shallow heating or accretion-induced heating. In these simulations, we do not model the mechanism producing the heating effect, just the effect itself, so we do not distinguish between which of these mechanisms cause the heating.

\begin{figure}[t]
\centering
\plotone{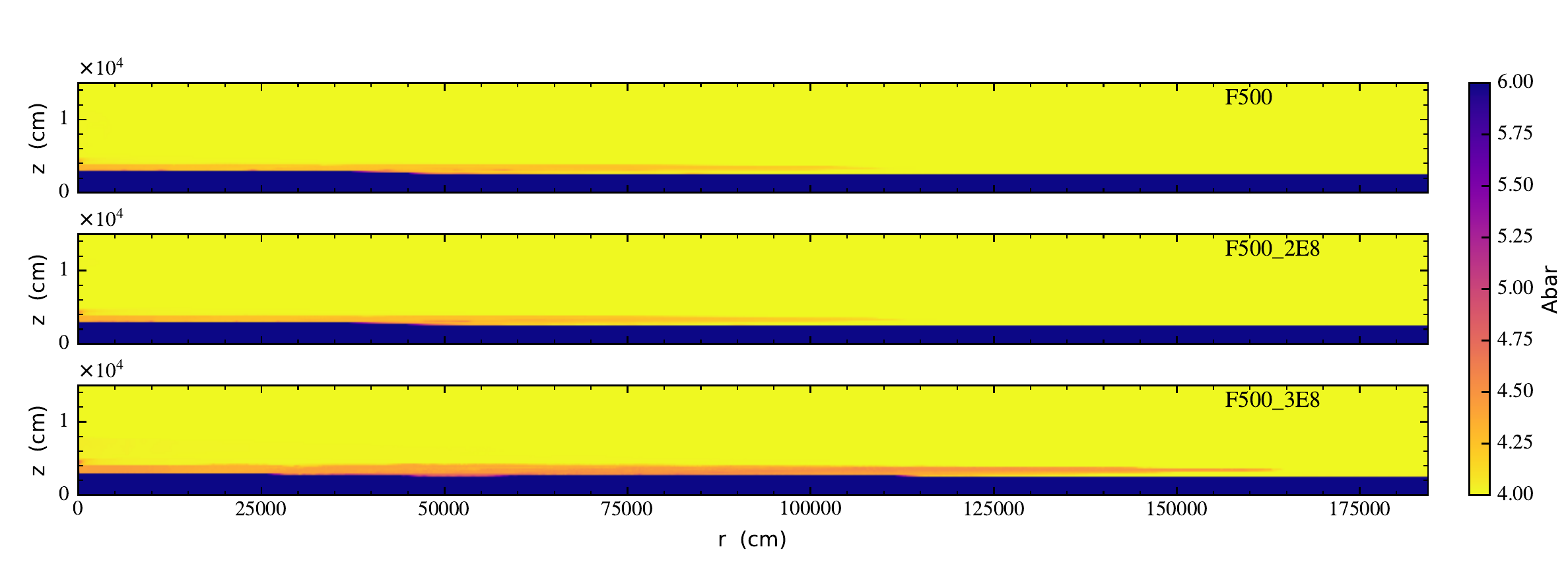}
\caption{\label{fig:compare_500Hz_abar} Comparison of $\bar{A}$ for 3 different $500~\mathrm{Hz}$ models with neutron star temperatures $T_{\rm star}$ of $10^8~\mathrm{K}$, $2\times10^8~\mathrm{K}$ and $3\times10^8~\mathrm{K}$ ($\tt{F500}$, $\tt{F500\_2e8}$, and $\tt{F500\_3e8}$, respectively), and resulting envelope structures.  Each flame is shown at $70~\mathrm{ms}$.} 
\end{figure}

\begin{figure}[t]
    \centering
    \plotone{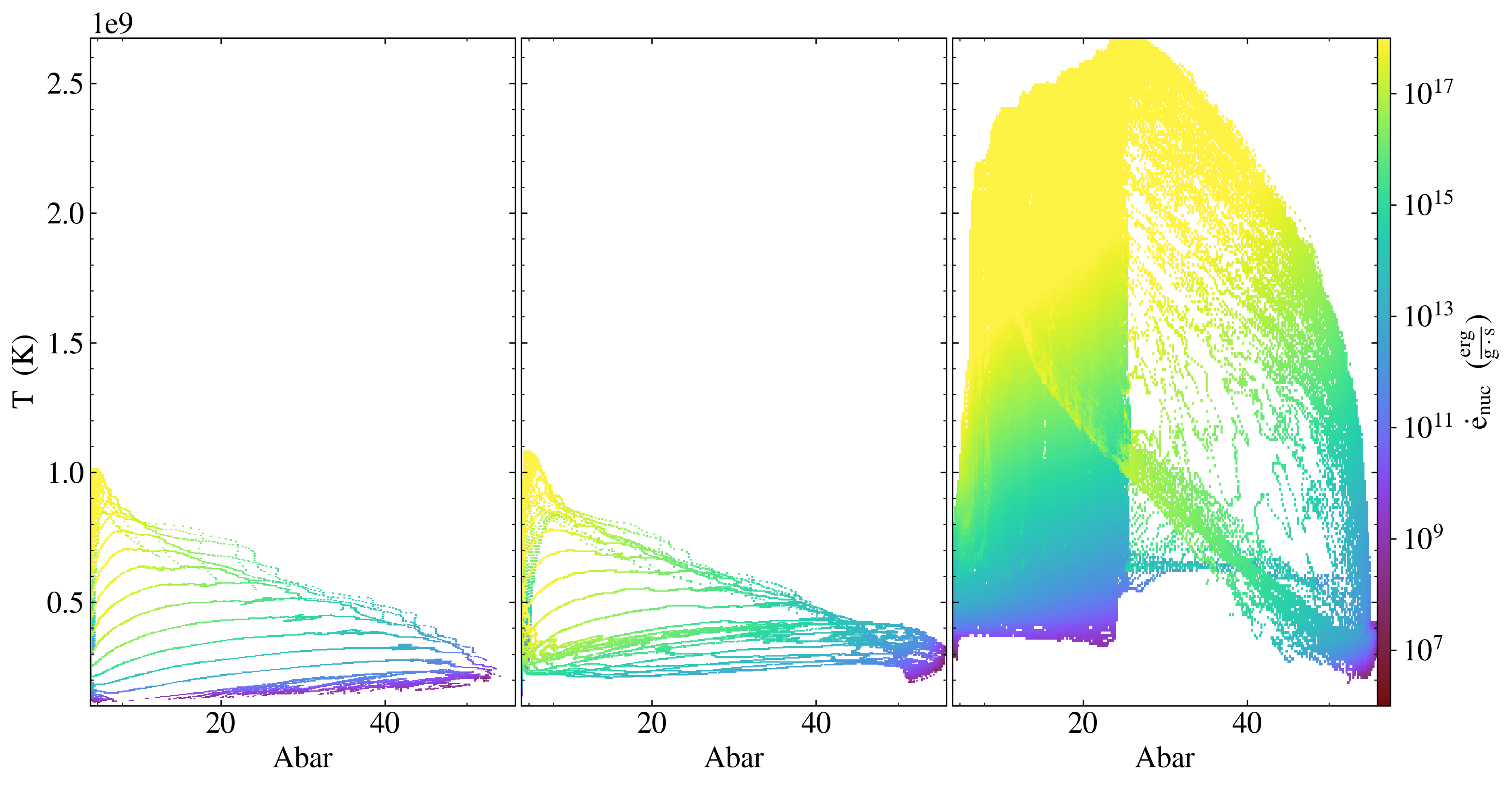}
    \caption{\label{fig:abar_hot}Phase plot of the mean molecular weight $\bar{A}$, temperature $T$ and the nuclear energy generation rate $\dot{e}_{\rm nuc}$ for the $500~\mathrm{Hz}$ runs with $T_{\rm star}$ of $10^8~\mathrm{K}$, $3\times10^8~\mathrm{K}$ and $4\times10^8~\mathrm{K}$ ($\tt{F500}$, $\tt{F500\_3e8}$, and $\tt{F500\_4e8}$) at $t = 40~\mathrm{ms}$. The run with $T_{\rm star} = 2\times10^8~\mathrm{K}$ ($\tt{F500\_2E8}$) closely resembles the cooler $\tt{F500}$ run, and is not shown here.}
\end{figure}

Figure~\ref{fig:compare_500Hz_abar} shows $\bar{A}$ for three $500~\mathrm{Hz}$ simulations with 
different initial temperature structures ($T_{\mathrm{star}} \leq 3 \times 10^8~\mathrm{K}$) at 
$t = 70~\mathrm{ms}$. We do not plot $\tt{F500\_4E8}$ here because it fails to form a clear burning 
front. $\tt{F500\_3E8}$ (Figure~\ref{fig:compare_500Hz_abar}, bottom panel) --- the hottest run to 
form a clear burning front --- has a faster propagating flame (this will be discussed further in 
Section \ref{ssec:temp_prop}). It also reaches slightly higher $\bar{A}$ values than the two cooler 
runs. The Figure~\ref{fig:abar_hot} $\bar{A}$-$T$ phase plots of $\tt{F500}$ (left) and 
$\tt{F500\_3E8}$ (middle) also reflect these $\bar{A}$ features, with $\tt{F500\_3E8}$ reaching 
slightly higher $\bar{A}$ values. $\tt{F500\_3E8}$ also reaches higher $\enucdot$ values at the 
low end of the temperature range ($< 0.5 \times 10^9~\mathrm{K}$). There appear to be more causally 
connected regions across a range of $\bar{A}$ at low temperatures for $\tt{F500\_3E8}$ than for 
$\tt{F500}$, suggesting that the higher $\enucdot$ for $\tt{F500\_3E8}$ generates burning in certain 
burning trajectories that are not present in the cooler $\tt{F500}$ run. Note that, although 
$\tt{F500\_3E8}$ is the hottest run with a modified initial temperature configuration to form 
a distinct flame front, the highest rotation $\tt{F1000}$ flame actually reaches higher temperatures 
(see Figure~\ref{fig:maxima}, left panel) as well as higher $\bar{A}$ (see Figure~\ref{fig:time_series_1000}).

\begin{figure}[t]
	\centering
	\plotone{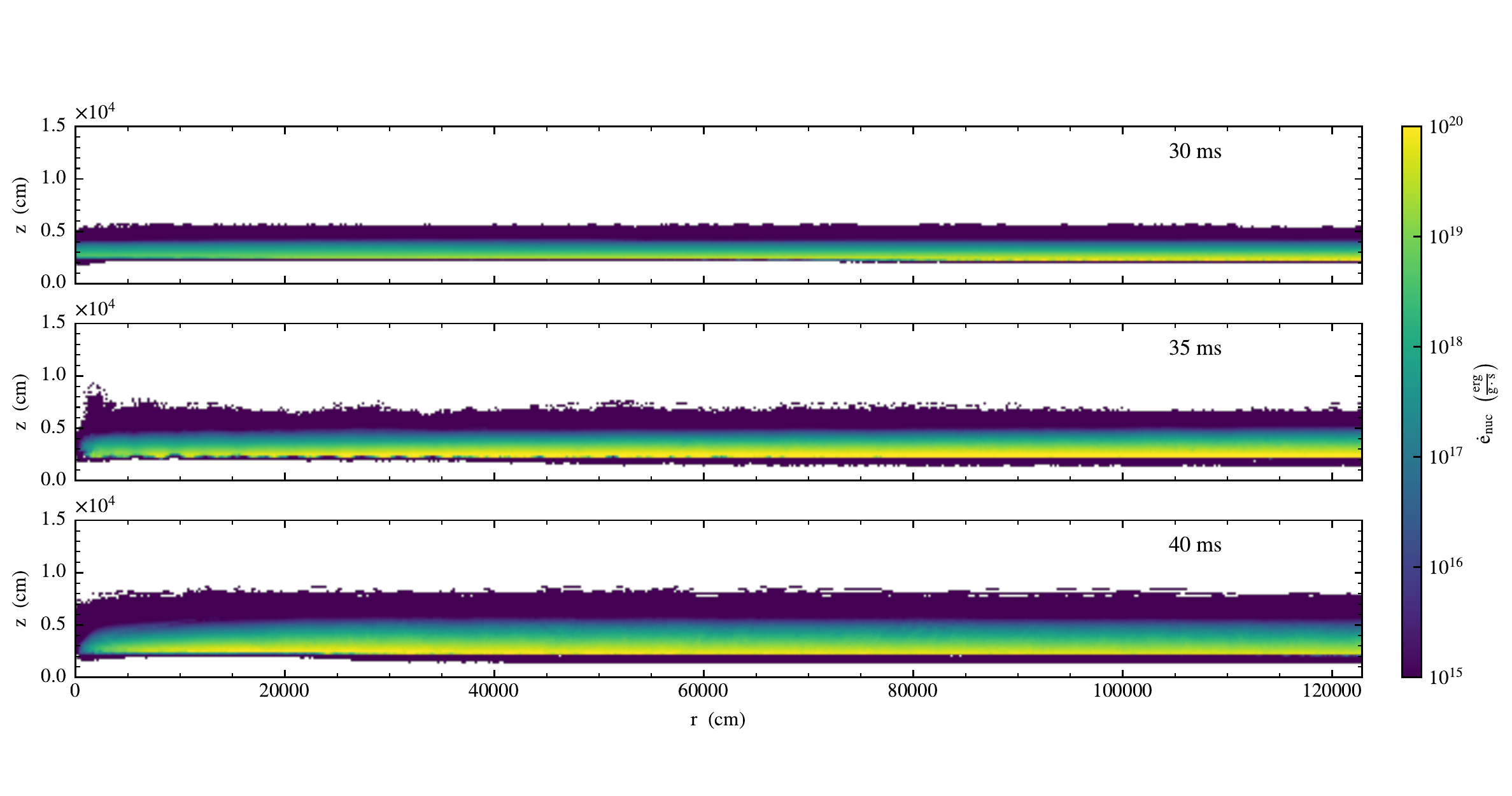}
	\caption{\label{fig:4e8_stacked_enuc} Time series of the $500~\mathrm{Hz}$ run $\tt{F500\_4E8}$, with $T_{\mathrm{star}} = 4 \times 10^8~\mathrm{K}$, showing $\enucdot$. This model burns so strongly that it is only run out to $40~\mathrm{ms}$; the snapshots shown here are at $T = 30~\mathrm{ms}$, $35~\mathrm{ms}$, and $40~\mathrm{ms}$.}
\end{figure}

\begin{figure}[t]
	\centering
	\plotone{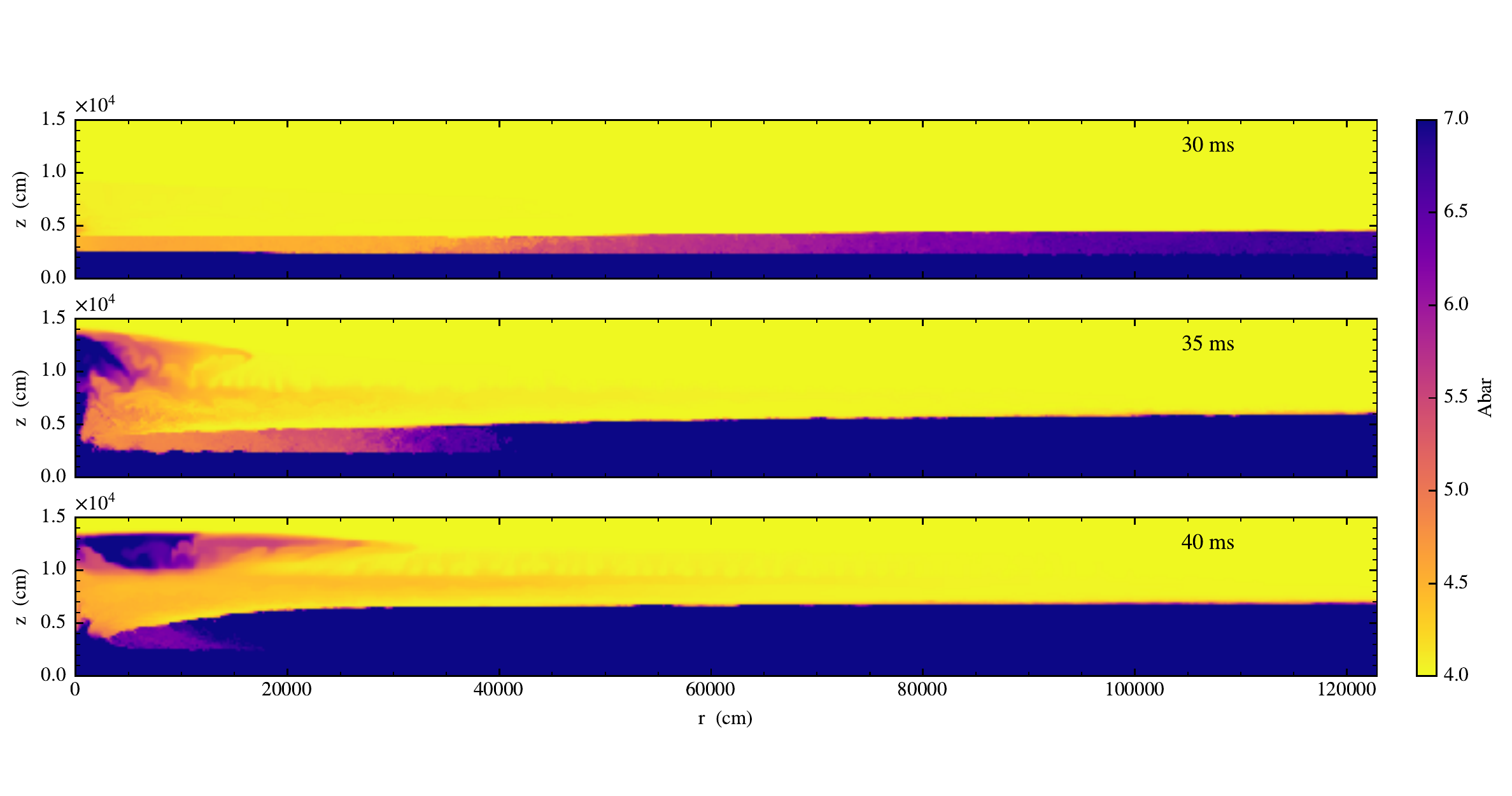}
	\caption{\label{fig:4e8_stacked_abar} Time series of the $500~\mathrm{Hz}$ run $\tt{F500\_4E8}$, with $T_{\mathrm{star}} = 4 \times 10^8~\mathrm{K}$, showing $\bar{A}$.}
\end{figure}

In contrast to the models with $T_{\mathrm{star}} \leq 3 \times 10^8~\mathrm{K}$, $\tt{F500\_4E8}$ 
is so hot that the organized flame structure is lost. \deleted{The fuel layer undergoes stable burning and 
essentially simmers in place, rather than forming a flame. This process has been observed in 
neutron stars with accretion rates that exceed the Eddington limit 
\citep{fujimoto1981,bildsten1998thermonuclear,keek2009effect}; in this model we have essentially 
recreated the conditions found in such stars with high accretion rates.} This model burns so 
strongly that it is only run out to $40~\mathrm{ms}$. After an initial period when the burning 
moves across the domain, residual burning continues and eventually ignites the entire fuel layer at 
late times, as shown in Figures \ref{fig:4e8_stacked_enuc} and \ref{fig:4e8_stacked_abar} for three 
snapshots taken in the last ten seconds of the simulation. In Figure~\ref{fig:4e8_stacked_enuc}, 
$\enucdot$ reaches values of $10^{18} - 10^{20}~\mathrm{erg}\,\mathrm{g}^{-1} \mathrm{s}^{-1}$ 
across the domain and at heights up to $\sim 0.5\times 10^4~\mathrm{cm}$. There is still 
significant burning occurring even higher, with $\enucdot$ reaching $\sim 
10^{15}~\mathrm{erg}\,\mathrm{g}^{-1} \mathrm{s}^{-1}$ at heights up to $\sim 0.8\times 
10^4~\mathrm{cm}$. For comparison, the next hottest run ($\tt{F500\_3E8}$) only reaches maximum 
$\enucdot$ values on the order of $10^{18}~\mathrm{erg}\,\mathrm{g}^{-1} \mathrm{s}^{-1}$ 
(see Figure~\ref{fig:maxima}, right panel), even at the latest timesteps ($\sim 
70~\mathrm{ms}$, when the flame is most developed). Significant $\enucdot$ values for 
all runs other than $\tt{F500\_4E8}$ are confined to the flame front, rather than spanning the entire domain.

\added{Rather than the fuel igniting in a 
localized hot spot and forming a flame which propagates across the domain, for this system the heat produced 
by the high temperature crust alone is sufficient to ignite material across the entire domain. Uniform burning 
across a large fraction of or even the entire surface of the neutron star has been observed in neutron stars 
with accretion rates that exceed the Eddington limit 
\citep{fujimoto1981,bildsten1998thermonuclear,keek2009effect}. However, it is important to note here that the
conditions in our model vary from the conditions that would occur in a neutron star ocean undergoing stable burning 
in a number of ways. Firstly, the model begins without any burning taking place, and the initial model assumes that a 
significant amount of unburnt fuel has been able to build up. For a stably burning system, freshly accreted material 
would be continuously burnt, so there would be much less unburnt fuel available. Secondly, we are only modelling 
a few tens of milliseconds, rather than an ongoing process. We therefore do not demonstrate that this system would 
continue to burn stably if the high crustal temperature were maintained and sufficient fuel continued to be 
available, e.g.~through the continued accretion of fresh material. However, this run does suggest that 
a crust with a sufficiently high temperature can provide enough energy to heat the fuel above it to the point of 
ignition, and it may be that if these conditions were maintained that it would be able to sustain stable burning.}

Figure~\ref{fig:4e8_stacked_abar} shows $\bar{A}$ for $\tt{F500\_4E8}$. Again, burning extends 
across the domain and high into the atmosphere and fuel layer, lacking the characteristic flame 
structure shown in $\bar{A}$ plots for the lower temperature $500~\mathrm{Hz}$ runs (see 
Figure~\ref{fig:compare_500Hz_abar}). A distinct mass of material appears to have broken out of the 
atmosphere near the axis of symmetry. The atmosphere edge is located at $\sim 1.2\times 
10^4~\mathrm{cm}$. A similar effect is also visible (to a lesser extent) in the $\tt{F500\_3E8}$ 
plot in Figure \ref{fig:compare_500Hz_abar}, with a faint haze of material rising above the flame near the axis of symmetry. 
However, this is likely to be a numerical artefact rather than a true physical effect, and a result 
of the boundary conditions imposed at the axis of symmetry for these simulations. $\tt{F500\_4E8}$ 
clearly reaches much higher $\bar{A}$ values across the domain (especially at the latest snapshot, 
$t = 40~\mathrm{ms}$) compared to all the other runs described in this paper (see Figures 
\ref{fig:time_series_500}, \ref{fig:time_series_1000} and \ref{fig:compare_500Hz_abar}).

 Though $\tt{F500\_4E8}$ does not form a distinct burning front, it does achieve greatly enhanced burning, as shown in the Figure~\ref{fig:abar_hot} $\bar{A}$-$T$ phase plot (right panel). Not only is $\tt{F500\_4E8}$ significantly hotter overall compared to all other runs, but there is a large causally connected region across a wide range of $\bar{A}$. This indicates that the hotter temperature has facilitated significant burning in burning trajectories that were not favored at the lower temperatures. The burning trajectories are also very disrupted for $\tt{F500\_4E8}$ compared to the cooler runs, suggesting that the hotter temperature leads to more vigorous mixing. Indeed, this appears to be the case looking at Figure~\ref{fig:4e8_stacked_abar} and the right panel of Figure~\ref{fig:uv_hot} (discussed further in section~\ref{ssec:temp_prop}). The disrupted burning trajectories resemble those found in Figure \ref{fig:abar} for the highest rotation $\tt{F1000}$ run, though they are even more dramatically disrupted for $\tt{F500\_4E8}$. $\tt{F500\_4E8}$ is also noticeably hotter than $\tt{F1000}$, even though it is run for significantly less time ($t = 40~\mathrm{ms}$ vs $t = 100~\mathrm{ms}$). Although the $\tt{F500\_4E8}$ run is clearly a special case in that it develops steady burning across the domain rather than a propagating flame, enhanced burning in this hottest run aligns with results from the other simulations with differing initial temperature structures. As $T_{\mathrm{star}}$ is increased, the overall behavior and propagation of the flame is significantly altered, implying that burning is very temperature-sensitive. We explore flame propagation further in Section~\ref{ssec:temp_prop}.

\subsection{Effect of Temperature on Flame Propagation}\label{ssec:temp_prop}

\begin{figure}[t]
	\centering
	\plotone{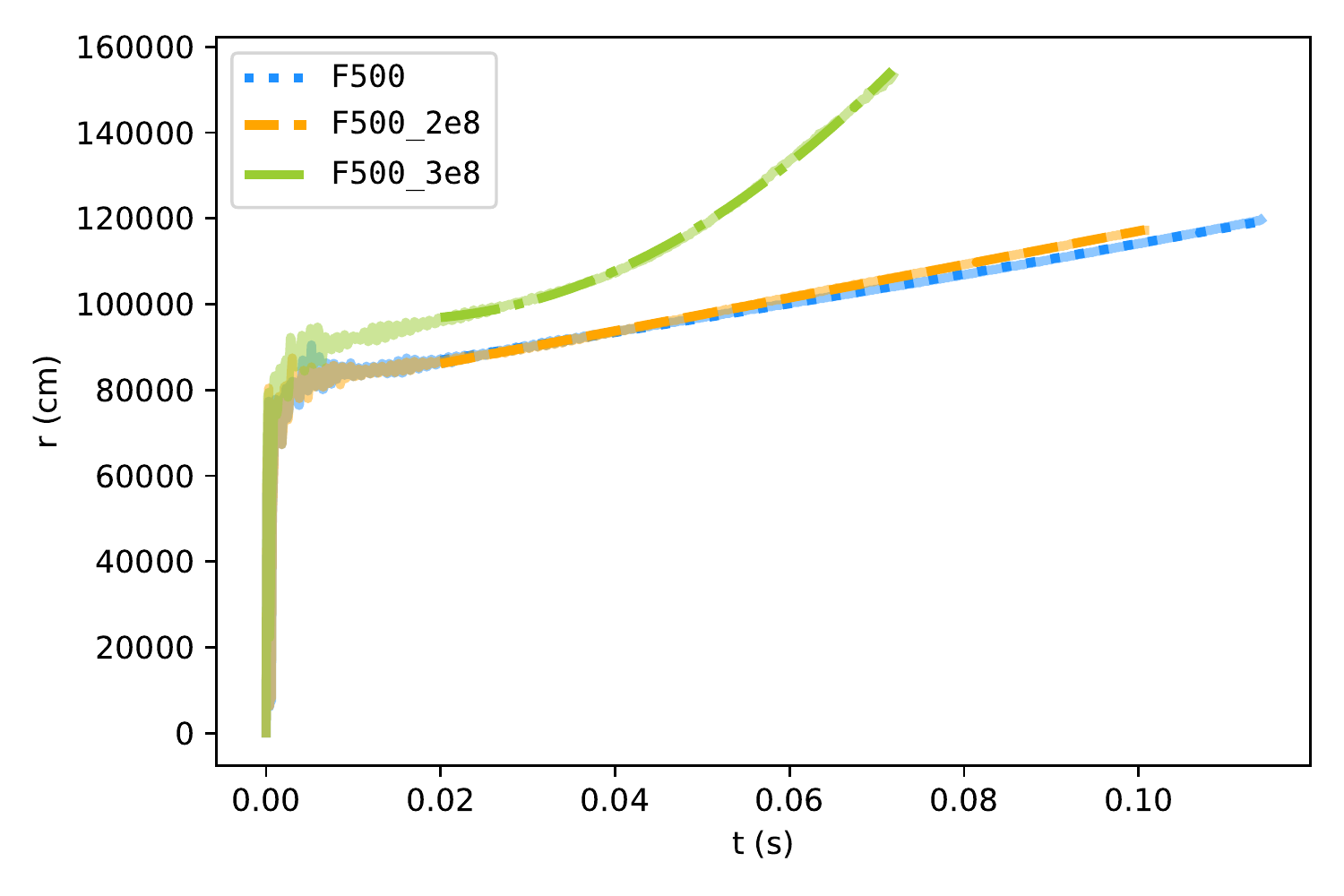}
	\caption{\label{fig:flame_speeds_2} Flame front position vs.\ time for the three $500~\mathrm{Hz}$ runs with
		$T_{\rm star}$ of $10^8~\mathrm{K}$, $2\times10^8~\mathrm{K}$ and $3\times10^8~\mathrm{K}$ ($\tt{F500}$, $\tt{F500\_2e8}$, and $\tt{F500\_3e8}$, respectively). The dashed lines show quadratic least-squares fits to the data
		for $t \gtrsim 20~\mathrm{ms}$. Note that due to its rapid acceleration, $\tt{F500\_3E8}$ 
		(green) is only run out to $\sim 70~\mathrm{ms}$ to avoid surpassing the domain boundary.}
\end{figure}

The method for measuring flame propagation described in Section \ref{ssec:rot_propagation} is 
applied in Figure \ref{fig:flame_speeds_2} to the three $500~\mathrm{Hz}$ runs with 
$T_{\mathrm{star}} \leq 3 \times 10^8~\mathrm{K}$. Due to the lack of a clear burning front in 
$\tt{F500\_4E8}$, we do not analyze its propagation velocity and acceleration. As the initial 
$T_{\mathrm{star}}$ is increased beyond $\sim 2\times 10^8~\mathrm{K}$, the flame becomes greatly accelerated. The initial flame 
velocities and accelerations derived from a quadratic least-squares fit to each of the datasets, as 
well as the flame velocities at $t = 50~\mathrm{ms}$ calculated using the fit parameters, are 
provided in Table~\ref{table:flame_speeds}. Comparing the flame propagation at different initial 
temperatures, the most robust feature is the acceleration of the $\tt{F500\_3E8}$ run at $t \sim 
40~\mathrm{ms}$. The reason for the acceleration of the flame is not entirely clear. 
Whereas for the cooler runs, a state of steady flame propagation is achieved, for the $\tt{F500\_3E8}$ run
the flame speed continues to increase, suggesting that some instability driving the 
flame's propagation persists to later times. It could be 
that energy released from burning begins to dominate the flame's propagation as it evolves, 
increasing the flame speed over time. Another possibility is that the increased temperatures lead 
to enhanced turbulent mixing effects that pull in more fresh fuel for the flame to burn. Yet 
another possibility is that the higher initial $T_{\mathrm{star}}$ leads to a greater average 
temperature in the fuel layer over time, making it easier for the flame to burn the fuel and 
propagate.

\begin{figure}[t]
    \centering
    \plotone{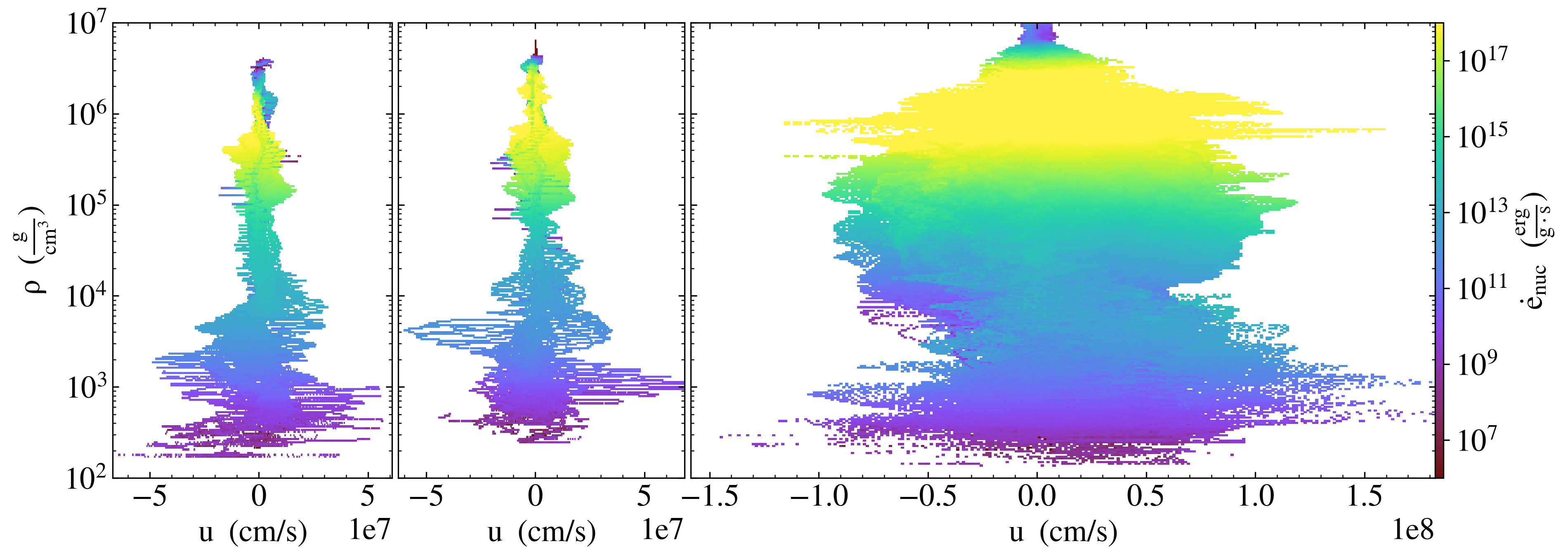}
    \caption{\label{fig:urho_hot}Phase plot of the horizontal $r$-velocity $u$, density $\rho$ and the nuclear energy generation rate $\dot{e}_{\rm nuc}$ for the $500~\mathrm{Hz}$ runs with $T_{\rm star}$ of $10^8~\mathrm{K}$, $3\times10^8~\mathrm{K}$ and $4\times10^8~\mathrm{K}$ ($\tt{F500}$, $\tt{F500\_3e8}$, and $\tt{F500\_4e8}$) at $t = 40~\mathrm{ms}$ (the latest time that $\tt{F500\_4E8}$ is run out to). The run with $T_{\rm star} = 2\times10^8~\mathrm{K}$ ($\tt{F500\_2E8}$) closely resembles the cooler $\tt{F500}$ run, and is not shown here.}
\end{figure}

\begin{figure}[t]
    \centering
    \plotone{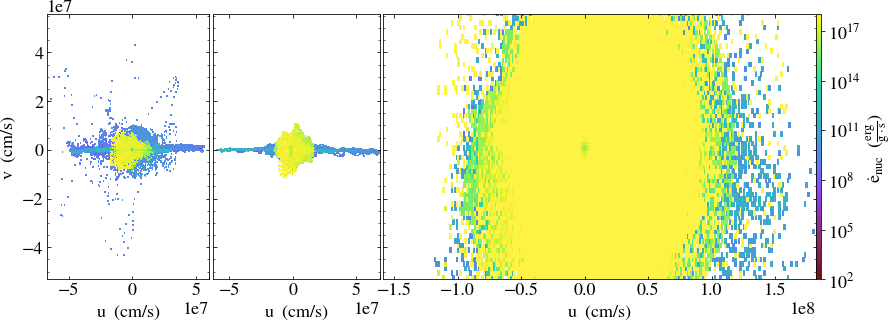}
    \caption{\label{fig:uv_hot}Phase plot of the horizontal $r$-velocity $u$, vertical $z$-velocity $v$ and the nuclear energy generation rate $\dot{e}_{\rm nuc}$ for the $500~\mathrm{Hz}$ runs with $T_{\rm star}$ of $10^8~\mathrm{K}$, $3\times10^8~\mathrm{K}$ and $4\times10^8~\mathrm{K}$ ($\tt{F500}$, $\tt{F500\_3e8}$, and $\tt{F500\_4e8}$) at $t = 40~\mathrm{ms}$. The run with $T_{\rm star} = 2\times10^8~\mathrm{K}$ ($\tt{F500\_2E8}$) closely resembles the cooler $\tt{F500}$ run, and is not shown here.}
\end{figure}

In both the $u$-$\rho$ phase plot in Figure~\ref{fig:urho_hot} and the $u$-$v$ phase plot in Figure~\ref{fig:uv_hot}, the horizontal $u$-velocity for the $\tt{F500\_4e8}$ (right panel) burning region is slightly larger than that of the cooler runs. Additionally, the cooler runs' $u$-velocities are two to three orders of magnitude greater than the flame speeds listed in Table~\ref{table:flame_speeds}. These phase plot velocities therefore most likely correspond to vortical motion in the turbulent burning vortices rather than the propagation of a flame itself. The vertical $z$-velocity $v$ in Figure~\ref{fig:uv_hot} further suggests that $\tt{F500\_4e8}$ undergoes increased vortical motion, as we see that the velocity magnitude in the burning region is significantly larger than in the lower temperature runs.
Also of interest in these plots is that the coolest run shows high velocity material comprising both high and low $\dot{e}_{\rm nuc}$ with horizontal velocity asymmetry, both of which suggest the burning vortex in the coolest case is less well defined than at higher temperatures. The coolest run thus does not seem to have developed the characteristic vortex structure at the burning front (i.e. where $\dot{e}_{\rm nuc}$ is greatest) that can be clearly seen for the hotter runs at this time ($40~\mathrm{ms}$).
As can be seen in Figure~\ref{fig:uv} (which was plotted at $100~\mathrm{ms}$), this does develop more at later times. Similar to what we saw when comparing the runs with different rotation rates, it would therefore appear that it takes longer for the flame to develop when $T_{\rm star}$ is cooler. 

\section{Discussion and Conclusions}\label{Sec:conclusions}

We ran a number of simulations of laterally propagating flames in XRBs in order 
to explore the effects of rotation and thermal structure. We found that increasing 
the rotation rate increased the energy generation rate within the flame and 
enhanced nuclear burning. Apart from the lowest rotation run (which failed to 
ignite), flame propagation was not noticeably impacted by rotation rate; by 
the time the different flame fronts reached steady propagation, they shared 
comparable velocities. These results are likely due to the rotation-dependent 
strength of the Coriolis force and its confinement of the flame balancing 
the enhanced nuclear burning. \added{This lack of dependence of the flame speed on 
the rotation rate could be tested using observations, as it implies that the burst 
rise time should have little dependence on the rotation rate. }

We explored several models with different 
crustal temperatures to determine what effects mechanisms such as higher 
accretion rates, crustal heating and shallow heating may have on flame 
propagation. We found that increasing the temperature of the crust 
significantly enhanced the flame propagation. This we believe to be 
because a cooler crust allows heat to more efficiently be transferred 
away from the flame itself, therefore reducing the flame's temperature, 
slowing burning and consequently reducing its propagation speed. At higher 
crustal temperatures, we saw that the inability for heat to be efficiently 
transported away from the flame front increased the flame temperature, 
driving unstable, accelerating flame propagation. We saw that if the crust 
temperature was too high, then instead of a flame the entire atmosphere 
would burn steadily. This is reminiscent of what is seen for neutron stars 
with accretion rates exceeding the Eddington limit. \added{This dependence of the 
flame speed on the crustal temperature could also be tested using observations, 
by looking to see how the burst rise time depends on the accretion rate (which 
may affect the crustal heating).}

In future work, we would like to improve and expand our
simulations in order to better understand the processes at play and to
include more physics. This includes adding tracer particles to the
simulations so we can monitor the fluid motion and perform more
detailed nucleosynthesis; extending our simulations to 3D, which would
hopefully alleviate some of the boundary effects we have observed in
these simulations but will require significant computational resources;
and exploring the resolution of our simulations more so that we can
ensure that we have resolved all of the necessary physical processes. We
would also like to model H/He flames, as these are the sites of
rp-process nucleosynthesis~\citep{rpprocess}.  Initially we will use the same reaction
sequence explored in our previous convection
calculations~\citep{xrb2}.  Finally, we
recently added an MHD solver to \castro~\citep{sazo2020thesis}; this
will allow us in the future to explore the effects of magnetic fields on flame
propagation in XRBs.



\acknowledgments \castro\ is open-source and freely available at
\url{http://github.com/AMReX-Astro/Castro}.  The problem setup used
here is available in the git repo as {\tt
  flame\_wave}.  \added{We thank the anonymous referee for their helpful suggestions.}
  The work at Stony Brook was supported by DOE/Office
of Nuclear Physics grant DE-FG02-87ER40317.  This material is based upon work supported by the
U.S. Department of Energy, Office of Science, Office of Advanced
Scientific Computing Research and Office of Nuclear Physics, Scientific
Discovery through Advanced Computing (SciDAC) program under Award
Number DE-SC0017955.  This research was supported by the Exascale Computing Project (17-SC-20-SC), a collaborative effort of the U.S. Department of Energy Office of Science and the National Nuclear Security Administration.
This material is also based upon work supported by the U.S. Department
of Energy, Office of Science, Office of Advanced Scientific Computing Research, Department of
Energy Computational Science Graduate Fellowship under Award Number DE-SC0021110. This work was supported in part by the U.S. Department of Energy, Office of Science, Office of Workforce Development for Teachers and Scientists (WDTS) under the Science Undergraduate Laboratory Internship (SULI) program. MZ acknowledges support from the Simons Foundation. 
This research used resources of the National Energy
Research Scientific Computing Center, a DOE Office of Science User
Facility supported by the Office of Science of the U.~S.\ Department
of Energy under Contract No.\ DE-AC02-05CH11231.  This research used
resources of the Oak Ridge Leadership Computing Facility at the Oak
Ridge National Laboratory, which is supported by the Office of Science
of the U.S. Department of Energy under Contract No. DE-AC05-00OR22725,
awarded through the DOE INCITE program.  We thank NVIDIA Corporation
for the donation of a Titan X and Titan V GPU through their academic
grant program.  This research has made use of NASA's Astrophysics Data
System Bibliographic Services. 

\facilities{NERSC, OLCF}

\software{AMReX \citep{amrex_joss},
          Castro \citep{castro,castro_joss},
          GCC (\url{https://gcc.gnu.org/}),
          Jupyter \citep{Kluyver2016},
          linux (\url{https://www.kernel.org/}),
          matplotlib (\citealt{Hunter:2007}, \url{http://matplotlib.org/}),
          NumPy \citep{numpy,numpy2},
          python (\url{https://www.python.org/}),
          valgrind \citep{valgrind},
          VODE \citep{vode},
          yt \citep{yt}}


\bibliographystyle{aasjournal}
\bibliography{ws}

\end{document}